\documentclass[aps,reprint,amsmath,amssymb,amsfonts,showpacs,nofootinbib]{revtex4-1}
\usepackage{graphicx}
\usepackage{mathrsfs}
\usepackage{color}
\usepackage{placeins}
\usepackage[T1]{fontenc}
\usepackage{bib_macro}
\usepackage[version=3]{mhchem}
\newcommand*{\fontpackage}{}
\renewcommand*{\fontpackage}{%
 txfonts%
}
\newcommand*{\fontpackageoptions}{}
\renewcommand*{\fontpackageoptions}{
}
\newcommand*{\isomathoptions}{}
\renewcommand*{\isomathoptions}{%
OMLmathsans,%
 sfdefault=zavm%     Arew sans (hab ich genommen) 
}
\usepackage[\fontpackageoptions]{\fontpackage}
\usepackage[\isomathoptions]{isomath}
\usepackage{esint}  %asukommentieren for mtpro2
\newcommand{\hc}{\mathrm{hc}}
\newcommand{\hor}{\mathrm{hor}}
\newcommand{\m}{\mathrm{m}}
\newcommand{\ma}{\mathrm{max}}
\newcommand{\su}{\mathrm{SU}}
\newcommand{\bh}{\mathrm{BH}}
\newcommand{\ti}{\mathrm{I}}
\newcommand{\tii}{\mathrm{II}}
\newcommand{\ini}{\mathrm{i}}
\newcommand{\ra}{\mathrm{r}}
\newcommand{\md}{d\hspace{-0.3mm}}
\newcommand{\hatt}{{\hat{t}}}

\newcommand{\hatc}{{\hat{\chi}}}

\newcommand{\bcdot}{\boldsymbol{\cdot}}

\renewcommand{\rho}{\varrho}
\newcommand{\ten}[1]{{\mbox{\footnotesize$\tensorsym{#1}$}}} %f�r kleine buchstaben 
\newcommand{\tenb}[1]{{\mbox{\small$\tensorsym{#1}$}}}  %f�r gro�e buchstaben 
\def\bigtiny{\fontsize{7pt}{7pt}\selectfont}
\newcommand{\tenin}[1]{{\mbox{\bigtiny$\tensorsym{#1}$}}}

\renewcommand{\vec}[1]{\hbox{{\boldmath$#1$}}}

\begin{document}
\title{Separate Universes Do Not Constrain Primordial Black Hole Formation}
\author{$^{1,2,3}$Michael Kopp}
\email{michael.kopp@physik.lmu.de}
\author{$^{1,2}$Stefan Hofmann}
\email{stefan.hofmann@physik.lmu.de}
\author{$^{1,3,4}$Jochen Weller} 
\email{jochen.weller@usm.lmu.de}
\affiliation{$^{1}$Excellence Cluster Universe, Boltzmannstr. 2, 85748 Garching, Germany \\
$^{2}$Arnold Sommerfeld Center for Theoretical Physics,Ludwig-Maximilians-Universit\"at,   \\
Theresienstr. 37, 80333 Munich, Germany   \\
$^{3}$University Observatory, Ludwig-Maximillians University Munich,  \\ Scheinerstr. 1, 81679 Munich, Germany \\ 
$^{4}$Max-Planck-Institut f\"{u}r extraterrestrische Physik, \\ Giessenbachstrasse, 85748 Garching, Germany}
%\affiliation{
%Ludwig-Maximilians-Universit\"at M\"unchen}
%-----------------------------------------------------OK-----------------------------------------------------------------------------------------------

%-----------------------------------------------------------Abstract-----------------------------------------------------------------------------------
%-----------------------------------------------------OK-----------------------------------------------------------------------------------------------
\begin{abstract}
Carr and Hawking showed that the proper size of a spherical overdense region surrounded by a flat FRW universe cannot be arbitrarily large as otherwise the region would close up on itself and become a separate universe. From this result they derived a condition connecting size and density of the overdense region ensuring that it is part of our universe. Carr used this condition to obtain an upper bound for the density fluctuation amplitude with the property that for smaller amplitudes the formation of a primordial black hole is possible, while larger ones indicate a separate universe. In contrast, we find that the appearance of a maximum is not a consequence of avoiding separate universes but arises naturally from the geometry of the chosen slicing. Using instead of density a volume fluctuation variable reveals that a fluctuation is a separate universe iff this variable diverges on superhorizon scales. Hence Carr's and Hawking's condition does not pose a physical constraint on density fluctuations. The dynamics of primordial black hole formation with an initial curvature fluctuation amplitude larger than the one corresponding to the maximum density fluctuation amplitude was previously not considered in detail and so we compare it to the well-known case where the amplitude is smaller by presenting embedding and conformal diagrams of both types in dust spacetimes.
\end{abstract}

%-----------------------------------------------------OK-----------------------------------------------------------------------------------------------
\maketitle

%-----------------------------------------------------OK-----------------------------------------------------------------------------------------------
\section{Introduction}
Einstein's general relativity (GR) permits the fascinating possibility that matter and fields interact with geometry in such a way that compact regions form from which temporarily no information can escape. Furthermore, assuming reasonable properties of matter and fields, the appearance of such a closed horizon (classically) implies the existence of a curvature singularity in the interior and that information can never escape \cite{HE73}. These configurations are black holes (BHs).

In a spherically symmetric spacetime, which we assume throughout the paper, a horizon and hence a BH forms as soon as the areal radius $R$ of a region which determines the measurable surface area  $A= 4 \pi R^2$, decreases and becomes smaller than twice the enclosed mass $M$ \footnote{We use natural units: $c=G=1$}. Achieving this matter configuration requires strong compression of the material and in most circumstances pressure forces will overcome the gravitational forces preventing the BH formation.

A dying star is a well-known exception. Depending on its mass gravity can overcome the forces exerted by the neutrons in the collapsed stellar core. In this way the quantum physics of neutrons determines the minimal mass of a stellar BH to be larger than the Tolman-Oppenheimer-Volkoff limit, which is a few solar masses.

Another possibility is that BHs form in the very early universe from primordial density perturbations \cite{ZN66, H71,CH74,C75}. These BHs are called primordial black holes (PBHs) and are the subject of this paper (we do not consider the formation involving the collapse or collision of defects, see for instance \cite{CS82,H89, RKS00, K10}). The formation process is bottom-up like in the ordinary structure formation scenario since only perturbations smaller than the Hubble radius $R<H^{-1}$ can collapse. The maximum size of a PBH forming in the early universe is approximately given by the Hubble radius $H^{-1}_\hc$ at the time $t_\hc$ when the perturbation crosses the horizon (see Sec.\,\ref{derivnosu}). As a consequence, PBH sizes range in principle between the Planck length and today's Hubble radius. The actual mass spectrum of PBHs then depends on the spectrum of primordial density perturbations and the equation of state. Numerical simulations of PBH formation in a radiation filled universe \cite{NNP78, BH79, SS99, HS02, MMR05, PM07} confirmed Carr's estimate \cite{C75}, that the initial fluctuation amplitude of a density perturbation must be very high in order not to disperse but to form a PBH.

PBHs would open up windows into many interesting aspects of GR. For instance, the formation process in a radiation-filled universe is a critical phenomenon \cite{C93,NJ99,HS02,MMR05, MMP08}. The mass spectrum of PBHs strongly depends on the probability distribution of metric fluctuations \cite{C75, ZMN82, GLMS04}, which depends in turn on the mechanism that generates the primordial density perturbations \cite{INN94, GLW96, BP97, I98, Y98, Y99}, such that a (non-)observation of PBHs constrains these mechanisms. And most importantly, while Hawking radiation \cite{H74} of stellar BHs is much colder than the CMB and hence undetectable, PBHs could have already evaporated or evaporate at the current time \cite{C75,C76, BCL91}, which could give rise to a  gravitational wave background \cite{P76, BR04, AEG09}. Additionally, quantum gravity effects may alter the evaporation process at its final stage such that there could have remained a Planck mass remnant for each evaporated PBH \cite{M87,BCL92}, while PBH evaporation could have created the baryons \cite{BCKL91,DNN00,AM07}. Many more reasons for studying PBH and more references can be found in \cite{C05}.

While there is still no direct observation of a BH, there is convincing evidence for the existence of stellar BHs \cite{MR03} and supermassive BHs in the centers of galaxies \cite{R84, A93}, in particular our own Galaxy \cite{GETAG09,BLN09}. In contrast there is no trace of PBHs and their existence is strongly constrained by observations \cite{JGM09,CKSY10}. However it is still possible that both dark matter and supermassive BHs may be the results of PBH evaporation \cite{M87} and PBH clustering \cite{C06,KRS05}.

\begin{figure}[t]
\centering
\includegraphics[width=0.25\textwidth]{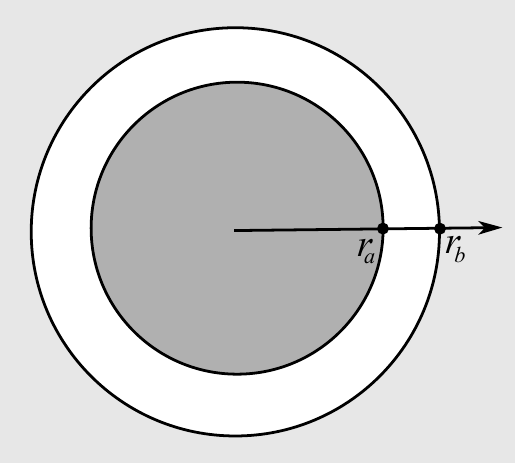}
\caption{Sketch of an idealized fluctuation: flat FRW for $r> r_b$ and closed FRW for $r<r_a$}
\label{sphconfig}
\end{figure}

The intention of the first part of this paper is to clarify a certain aspect of PBH formation which originated from the earliest treatments of PBHs \cite{H71, CH74,C75} and can still be found in recent works on PBH formation, e.g. in \cite{BP97,NJ98, HC05, MMP08, PM07, HP09} and probably many more. Carr and Hawking considered the collapse of an initially superhorizon-sized  homogeneous spherical overdensity surrounded by an otherwise flat FRW universe \cite{CH74,C75} (see Fig.\,\ref{sphconfig}). They estimated the necessary amplitude of the fluctuation for PBH formation using an energy argument: the gravitational energy at the moment $t_\m$ of maximal expansion, where the kinetic energy and the expansion rate is zero in the interior, has to overcompensate the internal energy density that will cause the pressure forces during the subsequent contraction. For a radiation-filled universe, they found that this requires the overdensity to have a size $R_\m \gtrsim \rho_\m^{-1/2}$, where $\rho_\m=\rho(t_\m)$ is the density inside the fluctuation at maximal expansion $t_\m$. On the other hand, it was noted that the space inside the fluctuation at $t_\m$ can only have positive curvature determined by $\rho_\m$. This leads to a maximal proper size $L_\m$ of the overdensity since the region would otherwise close up on itself and represent a separate universe (SU). They concluded that having no SU requires $R_\m \lesssim \rho_\m^{-1/2}$ \cite{CH74}. Hence Harada and Carr concluded that a fluctuation must be finely tuned in order to become a black hole but not a separate universe \cite{HC05}. One aim of this paper is to show that the no-SU condition actually does not pose a constraint on density fluctuations, and hence fine tuning is not necessary. We see this from several points of view in Secs. \ref{derivnosu} -- \ref{densfluc}. Carr's and Hawking's no-SU condition will be quickly rederived in Sec.\,\ref{derivnosu} keeping numerical factors in order to show why it does not constrain fluctuations. Defining the amplitude of a fluctuation in terms of a curvature fluctuation variable $\zeta$ in Sec.\,\ref{curvfluc}, we will conclude that a SU corresponds to $\zeta = \infty$ on superhorizon scales and hence SUs are physically impossible and irrelevant for PBH formation. In contrast to a result in \cite{C75}, we find in Sec.\,\ref{densfluc} that the density fluctuation $\delta$ features a maximal value $\delta_\ma$ which is independent of the no-SU condition and instead is a consequence of the spatial geometry of the fluctuation determined by the slicing and the Hamiltonian constraint. Furthermore it will be clear that this maximum arises naturally if one expresses $\delta$ in terms of $\zeta$. To every $\delta$ (except $\delta_\ma$) belong two different $\zeta$'s; one is larger and the other smaller than $\zeta_\ma\equiv\zeta(\delta_\ma)$. We will point out the relevance of the rather conceptual findings concerning SUs for the existing PBH literature in Sec.\,\ref{results}.
The second part of the paper is devoted to gaining intuition about PBH formation: in Sec.\,\ref{ltbcoll}, we compare the collapse of two spherical overdensities with the same $\delta$ but different $\zeta$ patched to a flat dust FRW universe using the LTB metric. The comparison will be carried out with the help of embedding diagrams and conformal diagrams, which enables us to intuitively understand the differences and similarities of both types of fluctuations. 

Throughout the paper we consider highly idealized models of density fluctuations. This is because they are analytically tractable and at the same time exhibit all the necessary features to clarify and show some new aspects of PBH formation. 

\section{Basic equations and no-SU condition}
\label{derivnosu}
In a spherically symmetric spacetime inhabited by a perfect fluid $\tenb{T}= (\rho + p)\, \ten{n}\! \otimes\! \ten{n} + p\, \ten{g}$ the metric $\ten{g}$ can be written in comoving coordinates, where $\ten{n}=1/N\,\ten{e}_{t},$ as
\begin{equation} 
\md s^2 =- N^2 \md t^2 +\frac{R'^2}{1+E} \md r^2  + R^2 \md\Omega^2.\label{metric}
\end{equation}
A prime denotes a derivative with respect to the comoving radial coordinate $r$ and $\md\Omega^2$ denotes the metric of the 2-sphere. Energy density $\rho$, pressure $p$, areal radius $R$, lapse $N$ and energy $E$ are in general functions of $r$ and $t$. Using metric \eqref{metric} in the Einstein equations  yields the so-called Misner-Sharp equations \cite{MS64}, which can be conveniently written using the Misner-Sharp mass
\begin{equation} 
M\equiv 4 \pi \int_0^r \rho R^2 R' \md r, \label{MSmass}
\end{equation}
that is the gravitating mass enclosed by $r$. In general, $M$ does not coincide with the integral of $\rho$ over the proper volume since
\begin{equation} 
\md V_\mathrm{B}\equiv4 \pi R^2 R' \md r\ \neq\  4 \pi R^2 R'(1+E)^{-1/2} \md r\equiv\md V_3. \label{voldiff}
\end{equation}
The Misner-Sharp equations are given by the Hamiltonian constraint,
\begin{equation} 
\dot{R}^2= N^2 \left(\frac{2M}{R}+E \right), \label{hamicons}
\end{equation}
the momentum constraint,
\begin{equation} 
\dot{E} = -2 \frac{1+E}{\rho +p}\frac{p'}{R'} \dot{R}, \label{momcons}
\end{equation}
the evolution equation,
\begin{equation} 
\dot{M}= -4 \pi p R^2 \dot{R}, \label{evoleq}
\end{equation}
and the Euler equation,
\begin{equation} 
(\ln N)' = - \frac{p'}{\rho+p}\;. \label{euleq}
\end{equation}
These equations are rederived in \cite{LL06} using the ADM formalism. 
In order to close the system of equations, we must provide an equation of state $f(\rho,p)=0$ and so as to solve \eqref{hamicons}--\eqref{euleq} we give initial conditions on an initial $t$=const-slice at $t_\ini$:
\begin{equation} 
R(t_\ini,r)=a_\ini r,\quad\rho(t_\ini,r)\quad\mathrm{and}\quad\dot{\rho}(t_\ini,r),
\end{equation}
where $a_\ini$ is a constant. This is only one possible choice for a set of initial conditions.

Consider now a spherical overdense region, whose homogeneous interior ($r<r_a$) is patched via a rarefied transition region to a flat FRW universe (see Fig.\,\ref{sphconfig}). We can then choose for $r> r_b$
\begin{equation} 
N= 1, \quad E=0\quad\mathrm{and}\quad R= b(t)\, r\;. \label{frwpart}
\end{equation}
Let us assume that the size of the fluctuation $R_\ini \gg H^{-1}_\ini$ is initially much larger than the Hubble radius of the flat FRW, where $H\equiv \dot{b}/b$, such that $p'\simeq 0$ and hence from \eqref{momcons} and \eqref{euleq} $E$ is approximately time independent and $N$ approximately constant\footnote{Note that the negligence of gradients is sometimes called ``separate universe approximation''. But this has nothing to do with the notion of a separate universe used in this paper.}. The interior then evolves like a closed FRW universe as long as this approximation is valid. This means that for $r<r_a$,
\begin{equation} 
N\simeq 1, \quad E\simeq -r^2\quad\mathrm{and}\quad R\simeq a(t)\, r\,. \label{flucapp}
\end{equation}
Requiring $p>-\rho/3$, the horizon $H^{-1}$ will grow faster than $R(t,r_a)$ and overtake it around the time $t_\mathrm{m}$ when the size of the overdense region becomes maximal. At this time $\dot{R}_\mathrm{m}$ vanishes and we get from \eqref{hamicons}
\begin{equation} 
E=-\frac{2M_\m}{R_\m}\;. \label{emaxexp}
\end{equation}
Since inside the fluctuation $\rho$ is spatially constant, Eq.\,\eqref{MSmass} gives $M=4\pi/3\,\rho\, R^3$. Inserting this together with \eqref{flucapp} into \eqref{emaxexp}, we obtain
\begin{equation}
16 \pi \rho_\mathrm{m} =\frac{6}{a_\mathrm{m}^2}. \label{rhomax}
\end{equation}
The right hand side is exactly the Ricci scalar \ce{^{(3)\!}$R$} of a 3-sphere with radius $a_\mathrm{m}$. The density of the flat FRW background $\bar{\rho}_\m\simeq \rho_\m$ will not deviate very much from the central density and from the Friedmann equation \eqref{frwpart} and \eqref{hamicons} it follows  that $H^{-1}_\m\simeq R_\m$, such that gradients become important and the approximation \eqref{flucapp} breaks down at least near the matching region when the fluctuation starts to collapse.
\begin{figure}[t]
\centering
\includegraphics[width=0.5\textwidth]{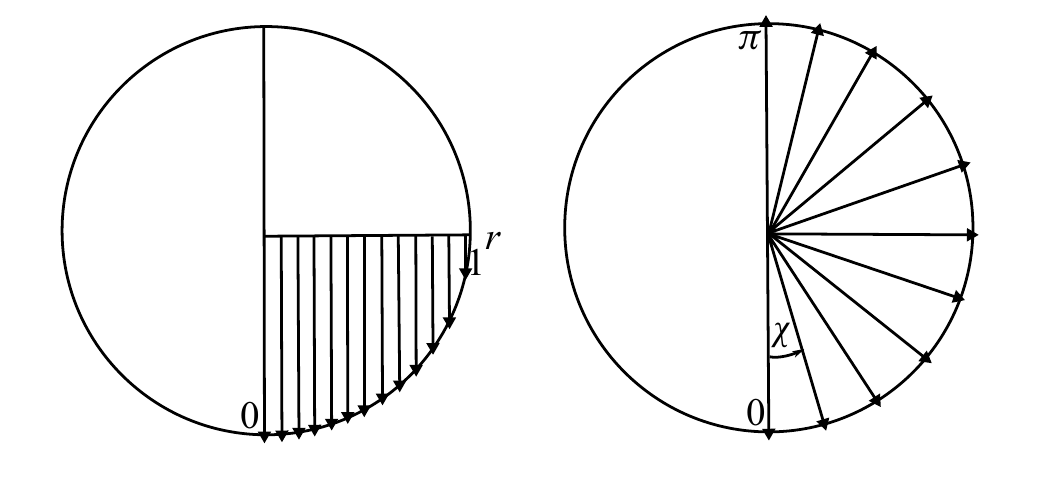}
\caption{Comparison of $r$ and $\chi$ radial coordinates. Both circles show an embedded 3-sphere ($\theta$ and $\phi$ directions suppressed).}
\label{radialcom}
\end{figure}
Let us introduce a new, approximately comoving\footnote{That $\chi$ is approximately comoving is a consequence of the approximate time independence of $E$.} radial coordinate $\chi$ defined by
\begin{equation}
\md \chi = \frac{\md r}{\sqrt{1+E}}\;. \label{chicoor}
\end{equation}
This change of coordinates will overcome any issues at points where $E=-1$ and enables us to treat ``normal'' PBH forming fluctuations $\zeta<\zeta_{\ma}$, the marginal $\delta=\delta_{\ma}$ and fluctuations with $\zeta>\zeta_{\ma}$  within the same coordinate system. For a closed FRW universe, where $E=-r^2$, introducing the $\chi$-coordinate solves the problem at $r=1$:
\begin{equation}
 \mathrm{\ce{^{(3)\!\!}$ds^2$}}=a^2\left(\frac{\md r^2}{1-r^2}+r^2 \md \Omega^2\right) = a^2\left(\md \chi^2+\sin^2 \!\chi \md \Omega^2\right)\,, \label{3metric}
 \end{equation}
which was just a coordinate singularity. The above metric is the metric of a 3-sphere with radius $a$, and $r$ parametrizes this 3-sphere by projecting down the areal radius $r$ of the equatorial unit ball (ranging from 0 to 1) and so the upper hemi-3-sphere is not covered (see Fig.\,\ref{radialcom}) . The new radial coordinate $\chi$ covers the whole 3-sphere and corresponds to its latitude (running from 0 to $\pi$).
\begin{figure*}[t]
\centering
\includegraphics[width=1\textwidth]{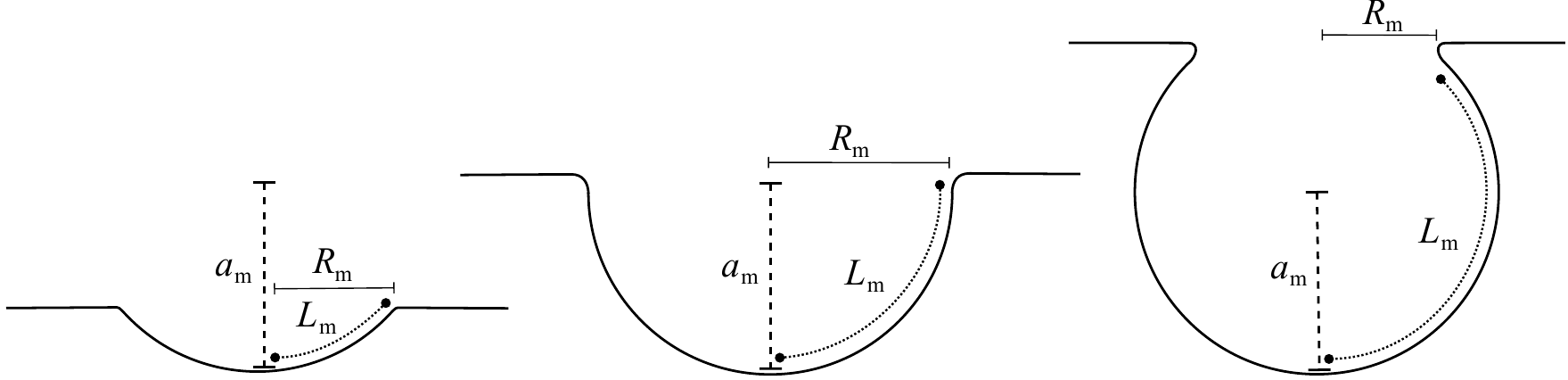}
\caption{Embeddings of spatial slices of different types of density fluctuations (angular dimensions suppressed). The left figure shows a case where $\zeta <  \zeta_\ma $ (, $\chi_a < \pi/2$, $\delta < \delta_\ma$) called type I, the middle figure shows the case  $\zeta =  \zeta_\ma $ (, $\chi_a =\pi/2$, $\delta = \delta_\ma$)  and the right one shows a case where $\zeta >  \zeta_\ma $ (, $\chi_a > \pi/2$, $\delta < \delta_\ma$) called type II.}
\label{supic}
\end{figure*}

From this we observe that the proper radius $L\equiv a \chi_a$ of the overdense region is constrained
\begin{equation}
L<L_\su \equiv a \pi\,. \label{nosuconGeo}
\end{equation}
Using the Hamiltonian constraint Eq.\,\eqref{rhomax} gives at $t_\m$
\begin{equation}
L_\su=\sqrt{\frac{3  \pi}{8 \rho_\mathrm{m}}}\;.\label{nosucond}
\end{equation}
An equality $L_\m=L_\su$ would mean that the assumed overdensity was in fact never part of our universe but instead a SU.
This no-SU condition ensures that there should be no matter configuration that is topologically impossible. Or in the words of Carr and Hawking \cite{CH74}\footnote{Their $\mu$ is our $\rho$.}:\\
\\
{\it The $R^{00}-\frac{1}{2} g^{00} R = 8 \pi T^{00}$ constraint equation implies that the 3-geometry of this hypersurface has positive curvature of order $\mu$ in the region where the rate of expansion is zero. If this positive curvature extended over a sufficiently large region, the spacelike hypersurface would close up on itself to form a disconnected compact 3-space of radius about $\mu^{-1/2}$. In this case the region would form a separate closed universe which was completely disconnected from our Universe. Such a situation would not correspond to a black hole formed by collapse of matter in our Universe. This shows that for black hole formation $\mu R^2 \lesssim 1$.} \\
\\
To obtain the estimate in the last line, which relates the areal radius $R_\m$  instead of the proper radius $L_\m$ to $\rho_\m$, we could argue that in a positively curved space 
\begin{equation}
R <  L \label{estimate}
\end{equation}
at $r\neq0$ and hence \eqref{nosuconGeo}
is violated if \eqref{rhomaxhaw} is violated.
\begin{equation}
R_\mathrm{m}<\sqrt{\frac{3  \pi}{8 \rho_\mathrm{m}}} \label{rhomaxhaw}
\end{equation}
Although our derivation of the no-SU condition \eqref{rhomaxhaw} seems to be correct and coincides with the cited result of \cite{CH74}, the last step \eqref{estimate} relating proper size $L$ and areal radius $R$ is a poor estimate. Looking at the embeddings of spatial slices at $t_\m$ for fixed $a_\m$ and different $L_\m$ in Fig.\,\ref{supic}, we see that the limit $L_\m \rightarrow L_\su$ actually requires $R_\mathrm{m}\rightarrow0$ instead of a violation of \eqref{rhomaxhaw}.
From the spherical geometry (see Fig.\,\ref{supic}) it is also clear that 
\begin{equation}
R<R_\mathrm{max} = a \label{geometry}
\end{equation}
is the maximum value $R_\m$ can take for fixed $a_\m$ and thus
\begin{equation}
R_\mathrm{m}\leq\sqrt{\frac{3}{8 \pi \rho_\mathrm{m} }}\;. \label{rhomaxhar}
\end{equation}
It is then obvious that the no-SU condition \eqref{rhomaxhaw} cannot be violated because the spatial geometry \eqref{geometry} of the overdensity ensures the stronger inequality \eqref{rhomaxhar}. It is important to note that there is a huge conceptual difference between the situations when an observable like $R_\mathrm{m}$ is constrained by avoiding a topology change \eqref{rhomaxhaw} or by simple geometry \eqref{rhomaxhar}. 

A further important remark is in order concerning the generality of our result. 
We used idealized fluctuations and neglected the pressure effects of the transition region because our argument is purely geometrical. It does not depend on Einstein's equations nor on the specific shape of the overdensity. In order to talk about a ``radius'' let us assume that the overdensity is approximately spherically symmetric, which should be a good approximation in case of a high amplitude fluctuation \cite{BBKS86}. For an inhomogeneously positively curved region \cite{ M41}, there exists an analog of inequality \eqref{nosuconGeo}, where $a$ is replaced by the maximal curvature radius of the region. However since \eqref{estimate} still strictly holds we see that $R$ 
is actually constrained by the areal radius of the ``equator'' (analog to Eq.\,\eqref{geometry}) of the positively curved region and not by the proper radius $L$ which is constrained by avoiding a SU.

Note that we would also obtain the bound \eqref{rhomaxhar} if we were to take the coordinate singularity at $r=1$ in metric \eqref{3metric} seriously for some reason. At this point $E=-1$ and  this was believed to indicate a SU \cite{H70,PM07}.
This may be the reason why type II fluctuations are excluded in the PBH forming parameter space in \cite{HP09} and why $\delta=\delta_\ma$ should indicate SUs \cite{C75}. We will see in Sec.\,\ref{densfluc} why $\delta_\ma$ corresponds to the upper bound in Eq.\,\eqref{rhomaxhar} and hence does not signal a SU. The appearance of a $\delta_\ma$ is related to the fact that the gravitating mass is given by $M \sim  V_\mathrm{B}$ instead of $M\sim V_3$ (see \eqref{MSmass} and \eqref{voldiff}). The existence of type II fluctuations ($\chi > \pi/2$) is  explicitly mentioned in \cite{BH79}, but not considered in their PBH formation simulations since these fluctuations should collapse anyway and are thus not interesting for calculating the PBH threshold $\chi_\bh$. 
Note also that for an asymptotic Minkowski space these configurations are known as semiclosed universes \cite{ZN83}.

Another reason why one could believe that $\delta=\delta_\ma$ marks the boundary to SUs is that fluctuations with $\zeta>\zeta_\ma$ actually become in a certain sense separate during their evolution on sub-Hubble scales. Following the time evolution of such a type II fluctuation in Sec.\,\ref{ltbcoll} we will observe that the neck that connects the bubble to the flat FRW gets stretched and tears itself apart. One might then be tempted to call the pinched off bubble a ``separate universe''. But as we will see in Sec.\,\ref{ltbcoll}, this fascinating process is simply an alternative mechanism for PBH formation. But it should be clear from the excerpt of \cite{CH74} and it was also pointed out in \cite{HC05} that a region can never evolve into a SU; it must be given as initial condition.
\section{Curvature fluctuation}
\label{curvfluc}
A physical mechanism that is at least in principle able to create type II fluctuations is single field inflation (see \cite{M05} for an introduction to inflation). To qualitatively see how this works, let us first define the curvature fluctuation $\zeta$ for a new radial coordinate $s$ that makes the spatial metric $\vec{\gamma}$ manifestly conformally flat:
\begin{equation}
\mathrm{\ce{^{(3)\!\!}$ds^2$}} = b^2(t) e^{2 \zeta(s,t)} \left(\md s^2  +   s^2 \md\Omega^2\right)\;. \label{ltbs2}
\end{equation}
The scale factor $b$ is the same as in \eqref{frwpart} if we set $r=s$ and $\zeta=0$ for $r>r_b$.

Roughly speaking inflation ends when the field $\phi$, which drives inflation with its potential $V$, reaches a certain value $\phi_\ra$ where both slow roll conditions break down. For the moment let us slice spacetime such that $\phi$ is spatially constant which makes in the  $\phi_\ra$-slice the reheating radiation energy density and also the expansion rate approximately constant. Nearly all the fluctuations are then in the local number of $e$-folds $$\lambda(s,t_\ra)\equiv\ln b(t_\ra)+\zeta(s,t_\ra),$$ which are directly related to fluctuations in the value of the local volume element $$\sqrt{\gamma(s,t_\ra)}=e^{3\lambda(s,t_\ra)} s^2 \sin \vartheta,$$ where $\gamma$ is the determinant of the spatial metric $\vec{\gamma}$ \eqref{ltbs2}. In this slicing the interpretation of how the quantum fluctuations of $\phi$ create the metric fluctuations is as follows. Regions where the inflaton took more $e$-folds to reach $\phi_\ra$ have a larger volume compared to the average. Let us estimate the volume fluctuation looking at Fig.\,\ref{supic}. It is determined by the ratio of the proper volume $V_3$ of the overdense region and the corresponding average flat space volume $V_\mathrm{B}$ one would measure if the fluctuation were absent. The proper volume is the volume of a 3-sphere cut off at a latitude $\chi$, using \eqref{voldiff} gives
\begin{equation}
V_3(\chi)= 2 \pi b^3 (\chi-\sin\chi \cos\chi), \label{vsph}
\end{equation}
which corresponds to $\sqrt{\gamma}$ integrated over the fluctuation. The volume of flat space is given by the ball at latitude $\chi$ is:
\begin{equation}
V_\mathrm{B}(\chi)=\frac{4\pi}{3}b^3\sin^3\chi\;.\label{vball}
\end{equation}
From this we get an averaged fluctuation $\bar{\zeta}$ defined by
\begin{equation}
\bar{\zeta}\equiv \frac{1}{3}\ln \frac{V_3(\chi_a)}{V_\mathrm{B}(\chi_a)}\;.
\end{equation}
For a marginal fluctuation with $\chi_a=\chi_\ma=\pi/2$ we obtain an average volume fluctuation $\bar{\zeta}_\ma\simeq 0.3$.
\begin{figure}[t]
\centering
\includegraphics[width=0.5\textwidth]{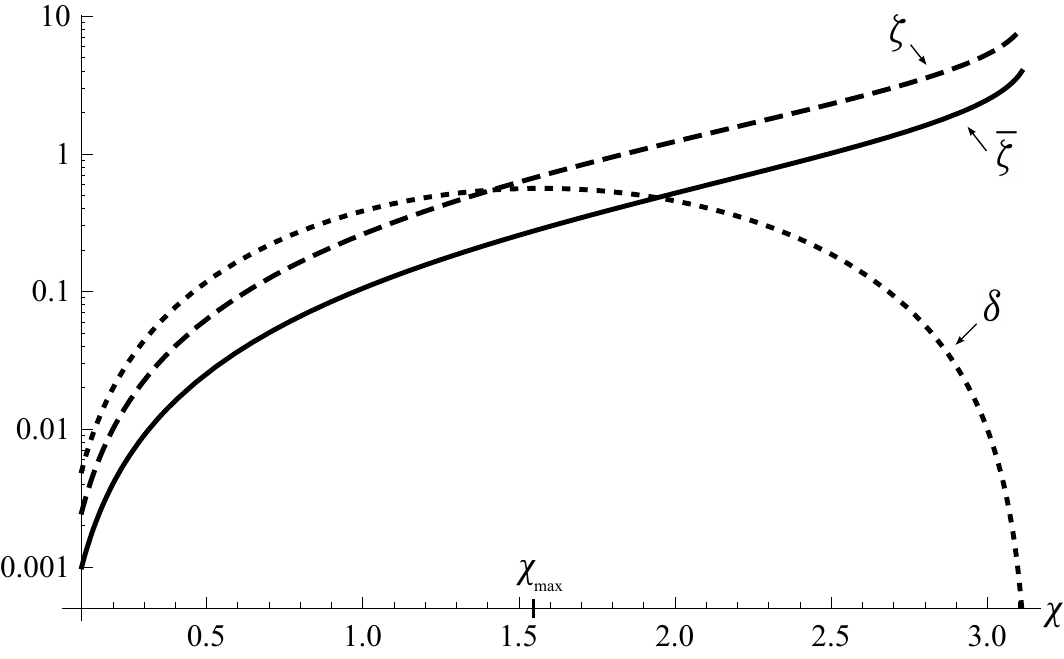}
\caption{Comparison of $\bar{\zeta}$, $\zeta$ and $\delta$}
\label{zetakom}
\end{figure}
We can compare this $\bar{\zeta}$ to the $\zeta$ defined by \eqref{ltbs2}.
In Appendix \ref{serfluc} we derive that as long as the size of the fluctuation exceeds the Hubble radius the central volume fluctuation $\zeta(0,t)$ is approximately given by:
\begin{equation}
    \zeta\simeq -2\ln \cos \frac{\chi_a}{2} \;. \label{zetaform2}
\end{equation}
Fig.\,\ref{zetakom} shows a comparison between $\zeta$ and $\bar{\zeta}$. The   fluctuation corresponding to $\delta_\ma$ is $\zeta_\ma\simeq 0.7$.

Both $\zeta$ and $\bar{\zeta}$ diverge for $\chi \rightarrow \pi$ and hence a fluctuation is a SU if and only if on superhorizon scales
\begin{equation}
 \zeta \;<\; \infty=\zeta_\su
\end{equation}
is violated. Obviously it is impossible to violate this no-SU condition.

It is also important to note that during matter or radiation domination, $\zeta$ will be approximately constant on super Hubble scales as long as spatial gradients are small (see eqs. \eqref{zetaformori} and \eqref{momcons} or \cite{LMS05}) and so the shape of the fluctuation will not change much. This implies that a type II fluctuation cannot be formed dynamically on super horizonscales during matter or radiation domination and  must be given as an initial condition on superhorizon scales.%: a type I superhorizon fluctuation (Fig.\,\ref{supic}) cannot evolve into a type II (or the other way around).
 
Inflation can produce these initial conditions at least in principle. In the desired comoving region the scalar field $\phi$ has to make quantum jumps up its potential hill a sufficient number of times such that inflation ends in that region on average 0.3 $e$-folds later. This should in principle be possible, although it may be very unlikely. However, this is a different question. All we want to emphasize is that the laws of physics allow the creation of arbitrarily large bubbles like the one on the right in Fig.\,\ref{supic}.

If inflation took place it created one large bubble connected to a mother universe by a comparably tiny neck: the large bubble is (or hosts) our universe while the neck has the size of the initial false vacuum region in the chaotic or phase transition undergoing preinflationary universe. Of course the curvature of this bubble does not have to be positive.

In Sec.\,\ref{ltbcoll} we will study the time evolution of a type II fluctuation: Once a neck gets smaller than the horizon of its surrounding mother universe it will start to stretch and eventually pinch off, separating the baby universe from the mother universe, leaving behind a black hole in both universes. Why and how this happens we will also see in Sec.\,\ref{ltbcoll}.  Hence every neck attached to a decelerating universe will sooner or later be destroyed. The only ambiguous sign left is the PBH. It is ambiguous because one can never figure out whether the PBH was formed by a type I or II fluctuation (Fig.\,\ref{supic}) since a type-I-PBH with a size $\chi_a<\pi/2$ has the same mass as a type-II-PBH with $\pi-\chi_a$ and hence they are indistinguishable for an observer in the mother universe. This ambiguity stems from the nonmonotonicity of $R(\chi)$. The density fluctuation variable $\delta$ which we will discuss next inherits this ambiguity too.

\section{Density fluctuation}
\label{densfluc}
In \cite{C75}, Carr derived from the no-SU condition in the form \eqref{rhomaxhaw} that there is a maximal density fluctuation $\delta_\ma$ with the property that for $\delta<\delta_\ma$ PBH formation is possible, while $\delta\geq\delta_\ma$ should indicate a SU. We know already that \eqref{rhomaxhaw} cannot be violated because of \eqref{rhomaxhar}. We will now show that $\delta_\ma$ is a consequence of \eqref{rhomaxhar} instead of \eqref{rhomaxhaw} and hence  reaching $\delta_\ma$ does not indicate a SU. Starting with the Hamiltonian constraint \eqref{hamicons} inside the fluctuation where $\rho$ is constant we get
\begin{equation}
\left(\frac{\dot{R}}{N}\right)^2= \frac{8 \pi}{3} \rho R^2+E\;. \label{hamiconsII}
\end{equation}
Like Harrison \cite{H70}, Carr \cite{C75} chooses an initial slice whose extrinsic curvature inside and outside the fluctuation is equal and constant, such that inside and outside the fluctuation
\begin{equation}
\frac{1}{N_\ini}\frac{\dot{R}_\ini}{R_\ini}=H_\ini\equiv \frac{\dot{b}_\ini}{b_\ini}\;.
\end{equation}
Inserting this together with $R=a_\ini r$ and $E=-r^2$ into \eqref{hamiconsII} gives
\begin{equation}
\frac{6}{a^2_\ini}+ 6H_\ini^2=16 \pi \rho_\ini\;. \label{hamiconshar}
\end{equation}
Outside the fluctuation, Eq.\,\eqref{hamicons} becomes at $t_\ini$ $$H_\ini^2=\frac{8\pi}{3}\bar{\rho}_\ini\;.$$ Dividing \eqref{hamiconshar} by $6 H^2_\ini$ gives
\begin{equation}
\left(\frac{H^{-1}_\ini}{a_\ini}\right)^2=\delta_\ini, \label{deltacons}
\end{equation}
with $$\delta(t)\equiv \Delta(t)-1 \qquad \Delta\equiv\frac{\rho}{\bar{\rho}}\;.$$
From the geometry of a 3-sphere (Fig.\,\ref{supic}) we have $R_\ini(\chi_a)\leq a_\ini$ which results in the following constraint for $\delta_\ini$:
\begin{equation}
\delta_\ini\leq \left(\frac{H^{-1}_\ini}{R_\ini(\chi_a)}\right)^2\equiv \delta_\ini^\ma\;. \label{deltaconsII}
\end{equation}
Fluctuations which are initially superhorizon-sized have to fulfill $\delta_\ini<1$ in this slicing. For fixed $H_\ini$ the density fluctuation $\delta_\ini$ hits this boundary for a size $\chi_a=\chi_\ma=\pi/2$ because $R_\ini(\chi_\ma)= a_\ini$. Hence fluctuations with amplitude $\zeta_\ma$ saturate the inequality, while fluctuations with $\chi<\chi_\ma$ and $\chi>\chi_\ma$ fulfill $\delta_\ini<\delta_\ini^\ma$, since $R_\ini(\chi_a)< a_\ini$.
To see how the no-SU condition $L_\ini<L_\su= a_\ini \pi$ constrains $\delta_\ini$ we can divide \eqref{deltacons} by $\pi^2$ and use the rather weak condition $R_\ini(r_a)<L_\su$ to get:
\begin{equation}
\delta_\ini<  \pi^2\delta_\ini^\ma\;. \label{deltaconsSU}
\end{equation}
The no-SU condition leads to a constraint \eqref{deltaconsSU} on $\delta_\ini$  which again is ensured by the geometry of space \eqref{deltaconsII}, such that the no-SU condition does not pose a constraint on $\delta_\ini$.

Because $\delta_\ini^\ma$ depends not only on the initial time $t_\ini$ but also on the initial size $R_\ini(r_a)$, it is more convenient to evaluate $\delta$ at the time of horizon crossing $t_\hc$, which is defined by  $R_\mathrm{hc}=H^{-1}_\mathrm{hc}$. In the following, we neglect gradients and pressure effects in the transition region between the closed and flat FRW part. We expect our expression for $\delta$ to be qualitatively correct, since only around the time $t_\hc$ this approximation starts to break down . 
Let's see what we find for
$$\delta\equiv \Delta(t_\mathrm{hc})-1$$
in the approximate synchronous gauge \eqref{frwpart} and \eqref{flucapp}, when we choose as initial condition a simultaneous big bang\footnote{Using the constant extrinsic curvature initial condition would lead to cumbersome expressions unless we would stick to small $\delta \ll 1$.}. To this end we use that the energy densities for closed and flat radiation-filled FRW universes are \cite{M05}:
\begin{equation}
\rho=\frac{3}{8\pi}\frac{1}{a_\mathrm{m}^2 \sin^4\eta} \qquad \bar{\rho}=\frac{3}{8\pi} (2 t)^2, \label{friedmann}
\end{equation}
where the conformal time $\eta$ ranging from $0<\eta<\pi$ is related to $t$ by
\begin{equation}
t= a_\mathrm{m}(1-\cos \eta)\;. \label{conftime}
\end{equation}
Any density fluctuation in the synchronous slicing has then $$\Delta_\mathrm{m}=4$$ at maximal expansion $\eta_\m= \pi/2$. For the marginal fluctuation $\chi_a=\chi_\ma=\pi/2$ with $$R_\mathrm{m}=a_\m=t_\m= \frac{1}{2} H^{-1}_\mathrm{m}$$ we see that it must reenter slightly before maximal expansion: $t_{\mathrm{hc}}<t_\mathrm{m}$, since $R_\m< H^{-1}_\m$. Because for fixed $\rho_\mathrm{m}$ the marginal fluctuation has the largest possible areal radius, types I and II will reenter earlier and we see already that $\delta$ features a maximum value $\delta_\ma$ and that it is again determined by the geometry and not SUs.
The calculation in Appendix \ref{derden} shows that
\begin{equation}
\delta\simeq\frac{1}{16}\sin^{2}\chi_a\left( 8+ \sin^{2}\chi_a \right), \label{delform}
\end{equation}
which is plotted in Fig.\,\ref{zetakom}. 
We see that the density fluctuation amplitude vanishes for a SU with $\zeta \rightarrow \infty$:
\begin{equation}
\delta_\su=\delta(\chi_a\!=\!\pi)=0\;. \label{deltasu}
\end{equation}
The interpretation of \eqref{deltasu} is clear: In the limiting case of a SU we have to close the bubble in Fig.\,\ref{supic}. We get two disconnected spacetimes, a flat and a closed FRW. The maximum value of \eqref{delform} is reached for $\zeta=\zeta_\ma$:
\begin{equation}
\delta_\ma=\delta(\chi_a\!=\!\pi/2)\simeq0.6\;.
\end{equation}
We used in this Sec.\,a radiation-filled spacetime as an example to show two important properties of $\delta$: the existence of a maximum independent of SUs and its vanishing approaching a SU. These properties hold in general because they follow from the behavior of $R$ in a positively curved space and the $R$ dependence of $M$ in Eq.\,\eqref{MSmass}. Since $\delta$ was defined at horizon entry, we only need to require that $k\equiv p/\rho>-1/3$. For a general $k>-1/3$ an expression for $\Delta_\mathrm{m}$ was calculated \cite{HC05}. 
\section{Consequences}
\label{results}
We briefly point out some consequences for the PBH literature regarding SUs. Since the findings are rather conceptual they will not change any numerical values found so far but they will shed new light on them.

Musco and Polnarev found $\tilde{\delta}_\ma\simeq 0.65$ for spherical fluctuation shapes including the rarefied transition region (see Fig. 11 in \cite{PM07}). They discuss that their $\tilde{\delta}$ is approximately $\delta$ and we see that the result from our simple model agrees quite well.

But we do not agree with the common physical picture of $\delta_\ma$, namely that it marks the boundary to SUs \cite{H70, C75,BP97,NJ98, HC05, MMP08, PM07, HP09}. Rather the appearance of $\delta_\ma$ is a consequence of (\emph{i})  the spatial geometry which depends on the chosen slicing and (\emph{ii}) the Einstein field equations, in particular the Hamiltonian constraint, which together always ensure the no-SU condition given in \cite{CH74}. The impossibility of its violation can be better understood in terms of the curvature fluctuation $\zeta$ Eq.\,\eqref{zetaform2}, where a SU is obtained as the limiting process $\zeta \rightarrow \infty$ during which the Hamiltonian constraint can be fulfilled. The utility of the no-SU condition given in \cite{CH74}, rederived as relation \eqref{rhomaxhaw} as well as the no-SU given in \cite{C75}, rederived as relation \eqref{deltaconsSU} is dubious since they have no geometric interpretation compared to relations \eqref{rhomaxhar} and \eqref{deltaconsII} that enforce the no-SU conditions anyway. It seems that we need not worry about SUs at all.

The expression for $\delta$ in Eq.\,\eqref{delform} is nonmonotonous in $\chi_a$ and hence $\delta$ is also nonmonotonous when written as a function of $\zeta$ using \eqref{zetaform2}. This means that physically distinct situations quantified by the volume fluctuation $\zeta$ can have the same $\delta$. This result is also valid for inhomogeneous and nonspherical density fluctuations. A general fluctuation may be considered as type II if the local areal radius $R$ becomes nonmonotonous. In the context of PBH formation this suggests not to use $\delta$ but instead to take $\zeta$ as fundamental for specifying the non-perturbative amplitude of a superhorizon sized fluctuation. Shibata and Sasaki did this in \cite{SS99} which automatically includes these type II fluctuations and avoids any coordinate singularities at $\delta_\ma$ which marks the boundary between types I and II.
%, while \cite{PM07} explicitely removed type II.

If one asks for the probability distribution function (PDF) of fluctuations it is also quite clear that the PDF for $\delta$ near $\delta_\ma$ can in no way be gaussian since it vanishes for $\delta > \delta_\ma$. The PDF for $\zeta$ could in principle be a gaussian near $\zeta_\ma$. But the initial condition for PBH formation from superhorizon-sized fluctuations is $\zeta \geq \zeta_\bh=0.7\div 1.2$ \cite{SS99}, which requires a nonperturbative method to obtain a reliable PDF, as done for instance in \cite{Y99}.

During the radiative era the PBH formation threshold $\delta_\bh$ is not far from $\delta_\ma$: numerical simulations \cite{PM07} found $\tilde{\delta}_\bh\simeq 0.5 \div 0.65$ for different fluctuation shapes. 
Depending on the PDF of the fluctuation random field $\zeta(\mathbf{x})$ from which one could derive a PDF for shape $S$ and amplitude $\zeta$ of fluctuations, the inclusion of type II fluctuations may be relevant concerning the PBH mass power law 
\begin{equation}
M_\bh = K(S)\cdot \left(\delta(\zeta)-\delta_\bh(S)\right)^{\gamma(S)} M_\hor
 \label{bhmasssc}
\end{equation}
for type I fluctuations determined in \cite{NJ99, MMP08}, where $M_\hor=4\pi/3 H_\hc^{-3}$ is the horizon mass of the unperturbed FRW at horizon crossing and $K$, $\gamma$ and $\delta_\bh$ depend on the shape $S$. The power law caused by the critical phenomenon diminishes the relevance of the most probable PBH forming fluctuations near $\delta_\bh$ for the PBH mass spectrum \cite{NJ98, GL99}. Even if the actual PDF renders type II fluctuations completely irrelevant, it is nonetheless interesting to note that the scaling law should change near $\delta_\ma$ to approximately
\begin{equation}
M_\bh\simeq\left(1+\delta(\zeta)\right) M_\hor \,.\label{bhmass}
\end{equation}
The mass is now determined by geometry and has the strange property that the PBH mass shrinks with increasing $\zeta$. We will comment on this at the end of the next section. One can find evidence for the behavior \eqref{bhmass}
 in \cite{MMP08}. We can extract this information from their Fig.\,1,
considering the last data point, which is closest  to $\delta_\mathrm{max}$. There we have approximately $\delta-\delta_\bh=0.1$. Taking the value $\tilde{\delta}_\bh \simeq\delta_\bh=0.5$ from \cite{PM07} (Fig.\,10 and $\alpha=0$), we get $\delta=0.6$.  Inserting this value into Eq.\,\eqref{bhmass} nearly coincides with $M_\bh=10^{0.2} M_\hor=1.6 M_\hor$ read off from the Fig. 1 in \cite{MMP08}. The authors mention that the scaling law starts to change at $\delta>0.01 +\delta_\bh$. And this change just seems to signal that the critical phenomenon ceases and that the ``normal'' horizon mass \eqref{bhmass} takes over near $\delta_\mathrm{max}$. 
\section{Collapse}
\label{ltbcoll}
We now know how to interpret $\delta_\ma$ correctly. To get a better understanding of what is actually going on during the collapse of a type II fluctuation, we will compare the collapse of a type II and type I fluctuation with the same $\delta$ but different $\zeta$. Instead of going through calculations we will observe the collapse using embedding diagrams, Figs. \ref{tgleich0} to \ref{tgleich25} and conformal diagrams, Fig. \ref{confopencloseII}. For calculational simplicity we take dust as the ideal fluid with $p=0$. For type I fluctuations note that if we choose $\delta$ large enough to escape the scaling law \eqref{bhmasssc}, the PBH formation process shown here will not qualitatively differ from the case where pressure is present. We will comment on the effect of radiation pressure for type II fluctuations in Fig.\,\ref{tgleich8}. Setting $p=0$ turns the Misner-Sharp system into that of LTB \cite{B47}. The Euler equation \eqref{euleq} allows us to choose $N=1$ everywhere. The momentum constraint \eqref{momcons} and the evolution equation \eqref{evoleq} make $E$ and $M$ time independent, such that using Eq.\,\eqref{MSmass}, the only degree of freedom is $R$. A solution $R(r,t)$ of \eqref{hamicons} with synchronous big bang is given by:
\begin{equation}
R(r,t)= \begin{cases} \displaystyle\frac{M}{E} (\cos \eta-1), \quad  \sin\eta-\eta=\frac{-|E|^{3/2}}{M} t &E(r)<0\\
\left(\displaystyle\frac{9M}{2}\right)^{1/3} t^{2/3} & E(r)=0
\end{cases},  \label{LTBsol}
\end{equation}
where $\eta$ is conformal time with $0<\eta<2 \pi$ and $t$ is cosmic time with $0<t<\infty$. We will not consider the case $E>0$.
\subsection{Initial data}
\begin{figure*}[t]
\centering
\includegraphics[width=0.32\textwidth]{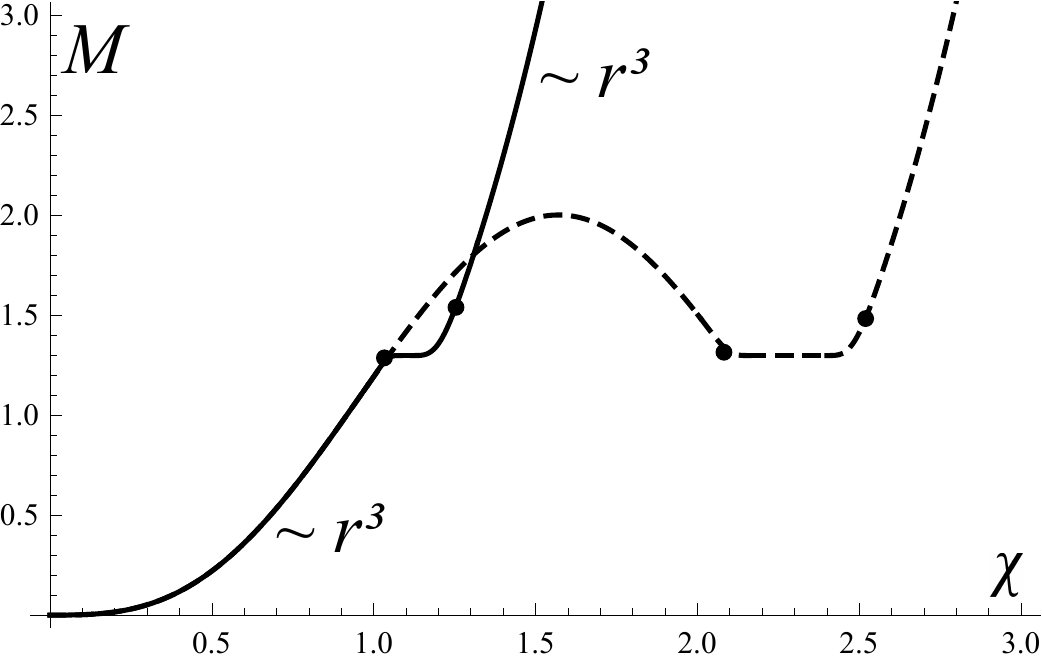}
\includegraphics[width=0.32\textwidth]{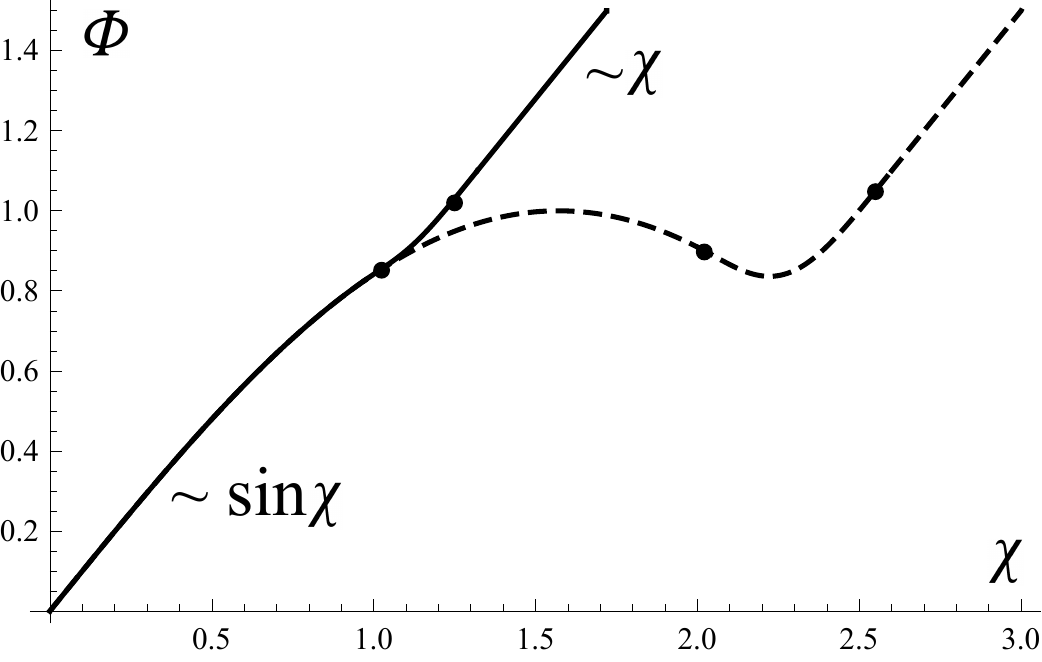}
\includegraphics[width=0.32\textwidth]{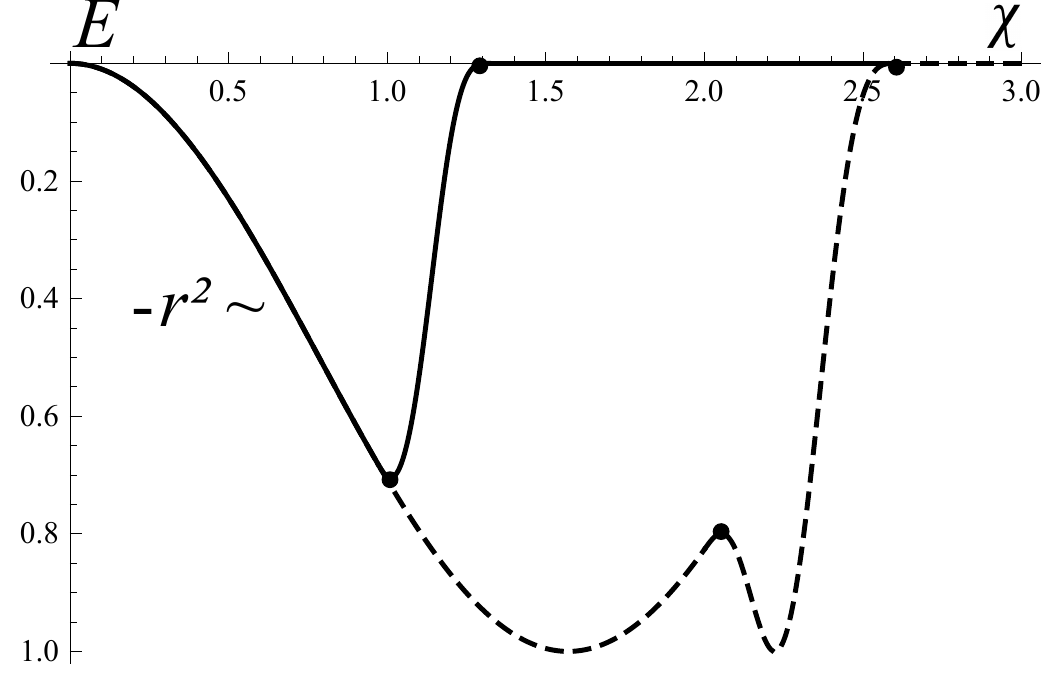}
\caption{Mass and energy functions. \emph{Full} lines correspond to type I and \emph{dashed} lines to type II}
\label{massenergy}
\end{figure*}

Because we have chosen $R(t_\ini\!=\!0,r)=0$, we cannot give $\rho_\ini(r)$ and $\dot{\rho}_\ini(r)$ as initial data, but we can equivalently give $E$ and $M$. In order to treat type II fluctuations without coordinate problems, we will use the now exactly comoving radial coordinate $\chi$ defined in Eq.\,\eqref{chicoor}. Writing $r(\chi)\equiv\Phi(\chi)$ and denoting a derivative w.r.t. $\chi$ by `, we have $$E(\chi)=\Phi`^2-1$$ such that we can give as initial condition for the simulations $$M(\chi)\quad \mathrm{and}\quad \Phi(\chi)$$ as depicted in Fig.\,\ref{massenergy}. They are chosen to match the following requirements:
\begin{itemize}
\item At every time conditions ({\it i}) - ({\it v}) given in Appendix \ref{LTBcons} are fulfilled.
\item The mass, size and formation time of both PBH types are equal. This just requires to choose for the sizes of the overdense regions to fulfill $\chi_\tii=\pi-\chi_\ti$. Note that $\chi_\tii$ and $\chi_\ti$ are time independent for dust.
\item The energy densities inside and outside the \emph{black} dots are homogeneous, while the transit region contains a matter free zone (the constant parts of $M$). Between the \emph{black} dots $M$ and $\Phi$ are sufficiently smooth interpolating functions.
\item $E\leq 0$ everywhere 
\end{itemize}
The first point guarantees a consistent and physically sane calculation. The second point makes comparison between both simulations easier. The third point is chosen to allow an analytic check of key events during the simulation. That there is a matter-free zone has only aesthetic reasons and does not influence much the dynamics of geometry. The last point ensures that the three-dimensional space can be embedded in an $\mathbb{R}^4$.

To really understand what the embedding diagrams show, let us briefly recapitulate the basic notions of embedding, intrinsic and extrinsic curvature, photon trajectories and horizons.
\subsection{Embedding and curvatures} 
\subsubsection{Embedding and intrinsic curvature}
At a fixed time the slice through spacetime is a Riemannian 3-manifold. Because the space is isotropic we can neglect the two angular coordinates $\vartheta$ and $\phi$ to facilitate the visualization of the geometry: we go to the equatorial plane $\vartheta= \pi/2$ and drop the $\phi$ direction. We can easily embed the remaining one-dimensional line into an $\mathbb{R}^2$ which then encodes all the intrinsic geometry of the full 3-D slice. Rotating around the symmetry axis would restore the $\phi$ direction. The graph can be found in the following way. Let us denote the auxiliary fourth spatial coordinate by $w$. The metric of the plane is given by
$$\mathrm{\ce{^{(2)\!\!}$ds^2$}}=\md w^2+\md R^2.$$
The induced Riemannian metric on the graph $w(R)$ should coincide with the original radial part of the LTB metric \eqref{ltbchi} because the angular part of both metrics $R^2 \md \Omega^2$ coincide:
$$\mathrm{\ce{^{(1)\!\!}$ds_\mathrm{ind}^2$}}= \left( \left(\frac{\md w}{\md R}\right)^2+1\right) \md R^2\stackrel{!}{=} \left(\frac{R`(\chi)}{\Phi`(\chi)}\right)^2 \md \chi^2=\mathrm{\ce{^{(1)\!\!}$ds_\mathrm{ori}^2$}}$$
This can be rewritten as
$$\left(\left( \frac{w`}{R`}\right)^2+1\right) R`^2= \left(\frac{R`(\chi)}{\Phi`(\chi)}\right)^2$$
from which follows
$$w`=|R`|\sqrt{\frac{1}{\Phi`^2}-1}.$$
This gives the $w$-component of the embedding:
$$w(\chi,t)=\int_0^\chi|R`(\chi ',t)|\sqrt{\frac{1}{\Phi`(\chi ')^2}-1} \, \md \chi '.$$
Note that this only works as long as the argument of the square root is non-negative, which is the case if $E \leq 0$. Slices or parts of slices where $E>0$ can only be embedded in Lorentzian auxiliary spaces.
The thick line in the following plots, the embedding, that can be thought of as the spatial slice itself is the graph $$(R(\chi,t), w(\chi,t))$$ parametrized by $\chi$.
The length of a segment of the graph between $\chi_1$ and $\chi_2$ is identical to the proper length of a radial line in the slice between $\chi_1$ and $\chi_2$. Also the intrinsic curvature \ce{^{(3)\!\!}$R$} of the original slice is just the curvature of the line (= 6/(radius of fitted circle)$^2$).
But intrinsic curvature is only half the truth:

\subsubsection{Extrinsic curvature}
The real slice is not embedded in a flat $\mathbb{R}^4$ but in a curved Lorentzian spacetime. And so there is more geometry hidden in the extrinsic curvature $\vec{K}=1/2\, \dot{\vec{\gamma}}$. Since the evolution of the LTB system is completely determined by the Hamiltonian constraint \eqref{hamicons},  which can be written as \cite{MTW73}
\begin{equation}
\ce{^{(3)\!\!}$R$}= K_{ij}K^{ij}-K^2+16 \pi \rho, \label{hamiconsltb}
\end{equation}
we see that if we were able to visualize $K_{ij}K^{ij}-K^2$ we would have visualized spacetime.
Fortunately this can be done. With $X\equiv R`/\Phi`$ we get from $$\gamma^{ij}=\mathrm{diag}(X^{-2},R^{-2},R^{-2}\sin^{-2}\vartheta) $$ and the extrinsic curvature $$K_{ij}= \mathrm{diag}(X\dot{X},R \dot{R}, R \dot{R} \sin^2\vartheta)$$ the slice (and embedding) intrinsic vector field
$$\vec{K}_{\!\chi}\equiv\vec{K}\bcdot  \ten{e}_{\!\hat{\chi}}=\frac{\dot{X}}{X}\ten{e}_{\!\hat{\chi}},$$
where $\ten{e}_{\!\hat{\chi}}$ is the unit vector in $\chi$-direction, and the rate of expansion of shells:
$$H_R\equiv \frac{\dot{R}}{R},$$ which are the amplitudes of $\vec{K}_{\phi}$ and $\vec{K}_{\vartheta}$. These functions completely determine the extrinsic curvature part of the Hamiltonian constraint:
\begin{equation}
K_{ij}K^{ij}-K^2=-4 H_R^2-2K^{\!\hat{\chi}} H_R. \label{extriLTB}
\end{equation}
\subsection{Photons}
To get a better grasp of time and causality, we will follow the way of a few radially moving photons. Their trajectories can easily be calculated from the null condition: a photon moves on geodesics whose tangent is always null. Hence a photon reaches every distance in zero proper time. For a radially moving photon this means:
$$0\stackrel{!}{=}-\md t^2+ \left(\frac{R`}{\Phi`}\right)^2 \md \chi^2\quad \Rightarrow\quad \md t=\pm  \frac{R`}{\Phi`}\md \chi, $$
where + applies for outgoing and -- for incoming photons. This equation is integrated and the result $\chi_\mathrm{ph}(t)$ is added to the diagram as colored point $(R(\chi_\mathrm{ph}(t),t), w(\chi_\mathrm{ph}(t),t))$.
\subsection{Horizons}
A PBH inevitably forms as soon as spacetime develops a closed region in which  outward traveling photons cannot increase their areal radius \cite{HE73}. The outermost boundary of this inner-trapped region is the BH apparent horizon. In a spherically symmetric spacetime this black hole apparent horizon is a centered 2-sphere within a spatial slice on which radially {\it outward} traveling photons have constant $R(\chi_\mathrm{BH})$. This means that the derivative of $R$ in this outward null direction $$\ten{v}\equiv \ten{e}_t+ \frac{\ten{e}_\chi}{X}= \ten{e}_\hatt+ \ten{e}_\hatc$$ vanishes: $$\partial_{\tenin{v}} R(\chi_\bh(t),t)=0.$$ Outgoing photons with slightly larger $r$ have as usual a growing $R$ along their radial geodesic, while photons with smaller $r$ decrease their $R$ along their radial outward-directed geodesic and are trapped. Another apparent horizon which we take as the outermost boundary of the outer-trapped region is the Hubble horizon on which radially {\it inward} traveling photons have constant $R(\chi_H)$. This means that the derivative of $R$ in this inward null direction $$\ten{u}\equiv \ten{e}_t- \frac{\ten{e}_\chi}{X}= \ten{e}_\hatt- \ten{e}_\hatc$$ vanishes: $$\partial_{\tenin{u}} R(\chi_H(t),t)=0.$$ Note that the notion of an apparent horizon is only local both in space and time: Consider an inward traveling photon at $\chi_\mathrm{ph}(t_1)>\chi_H(t_1)$ whose $R$ increases. Since the Hubble radius grows this photon can later at $t_2$ be inside $\chi_\mathrm{ph}(t_2)<r_H(t_2)$ and reach the center. We will see this happen in the simulation. For a BH horizon, as was mentioned in the Introduction, the rules are different: once trapped, always trapped.

Slightly rearranging equation \eqref{hamicons} we get:
\begin{eqnarray}
M&=&\nonumber\frac{R}{2}(\dot{R}^2-E)=\frac{R}{2}\left(\dot{R}^2-\frac{R'^2}{X^2}+1\right)\\
&=&\frac{R}{2}\left(\partial_{\tenin{v}} R\cdot\partial_{\tenin{u}} R+1\right). \label{masshoexp}
\end{eqnarray}
From this expression for the mass $M$ we can read off the nice result that a necessary condition for PBH formation is $R\leq 2M$ hence that matter must be within its Schwarzschild radius. At the Hubble radius this condition is also fulfilled, so we have to be a bit cautious. Inner-trapped regions have:
\begin{equation}
R \leq 2M \quad \mathrm{and} \quad \dot{R}<0. \label{BHhorizon}
\end{equation}Outer trapped regions have:
\begin{equation}
R \leq 2M \quad \mathrm{and} \quad \dot{R}>0. \label{Huhorizon}
\end{equation}
\subsection{The embedding diagrams}
From the following plots displaying the collapse of a type II (\emph{upper} panel) and type I fluctuation (\emph{lower} panel) in a LTB spacetime we see the following information:
\begin{itemize}
\item The \emph{thick} (\emph{black}) line is the embedded 3d-space of constant time $t$, where we keep only the radial direction. Rotating around the symmetry axis would restore the $\phi$-direction. Intrinsic curvature \ce{^{(3)\!\!}$R$} and proper length are directly visible.
\item Tangential arrows show the vector field $\vec{K}_{\!\chi}$ that measures how fast space contracts or expands in $\chi$-direction. The arrows are \emph{tangential} and \emph{outward-directed} (\emph{green}) if space expands and \emph{tangential} and \emph{inward-directed} (\emph{blue}) if space contracts in $\chi$-direction.
\item Arrows in $R$-direction symbolize amplitude and sign of $H_R$, which measures how fast $R$-shells contract or expand. The arrows are \emph{horizontal} and \emph{outward-directed} (\emph{orange}) if the shell expands and \emph{horizontal} and \emph{inward-directed} (\emph{red}) if the shell contracts. Note that the direction of the arrows only symbolizes this information. Since the auxiliary embedding space has no physical meaning those arrows do not show a vector field. The vectors $\vec{K}_{\phi}$ and $\vec{K}_{\vartheta}$ would be tangential to $R$-shells with amplitude $H_R$.
\item The \emph{thick} embedding line is colored \emph{orange} (\emph{grey} for a b/w print), if the region is outer-trapped, \emph{red} (\emph{dark grey} for a b/w print) if it is inner-trapped and \emph{black} otherwise. The outer boundary between \emph{orange} and \emph{black} segments corresponds to the Hubble radius while the outer boundary of \emph{red} and \emph{black} regions corresponds to a BH horizon.
\item  Beyond the outer \emph{small} (\emph{black}) dot the energy density is that of a flat FRW while inside the interior \emph{small} (\emph{black}) dot the energy density is that of closed FRW. Approximately the \emph{small} (\emph{black}) dots denote the boundary of the matter-free region in the transit region between flat and closed FRW.
\item To elucidate the causal structure we will follow the trajectories of three test photons which are symbolized by \emph{large} (\emph{colored}) dots.
\end{itemize}

\newpage
\FloatBarrier
\begin{figure}[!]
\includegraphics[width=0.49\textwidth]{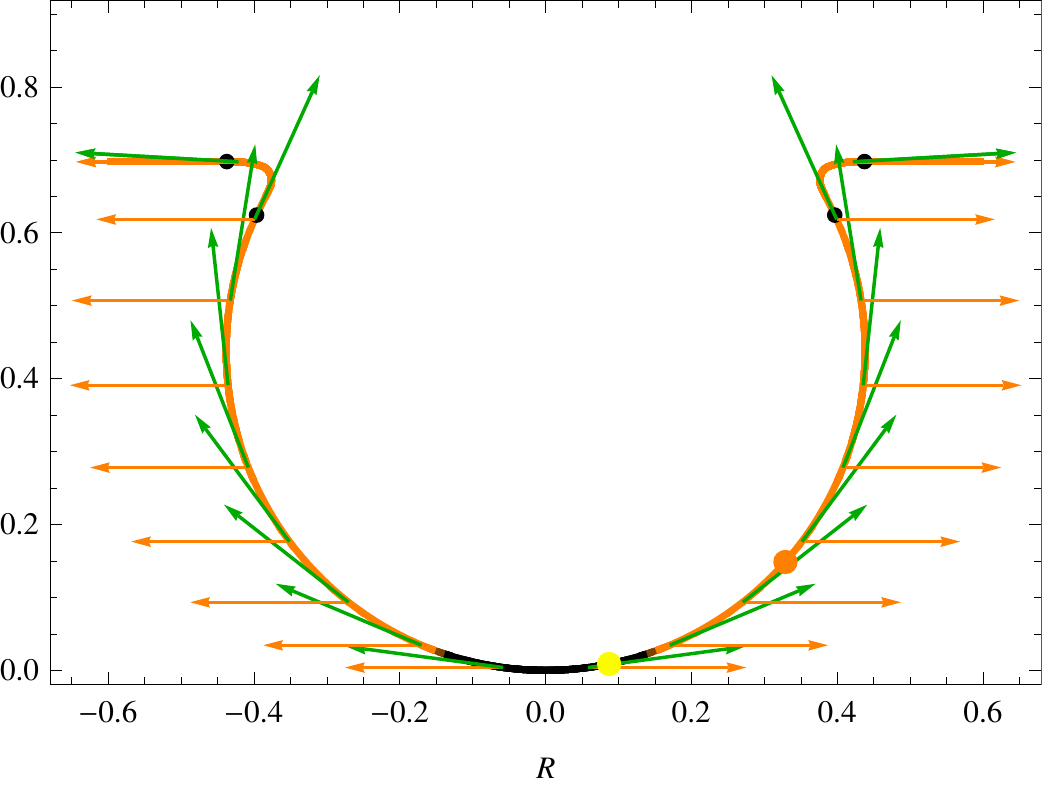}\newline
\includegraphics[width=0.49\textwidth]{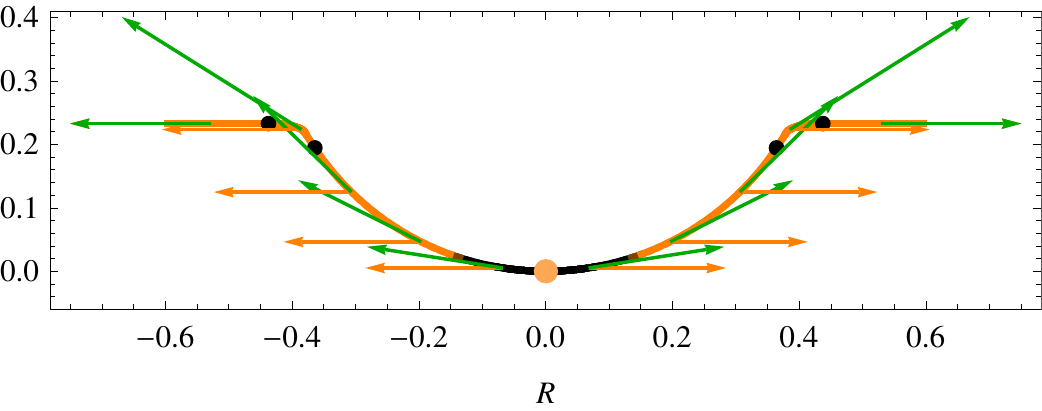} 
\caption{$t=0.1$\newline
Since the Hubble horizon is tiny (there is only a small \emph{black} region in the center) the spatial shapes are nearly unchanged at this early time compared to its initial values at $t_\ini=0$. Both extrinsic curvatures $K^{\!\hat{\chi}}$ and $H_R$ are very large (\emph{long} arrows) and positive (\emph{outward-directed} arrows) corresponding to the small Hubble radius. The arrows are actually 30 times longer and were rescaled to fit the paper. The following pictures however show the correct length.\newline
\emph{top}: Two outgoing photons, the \emph{yellow} and \emph{orange} dots (\emph{light grey} and \emph{grey} for a b/w print) start in the middle. \newline
\emph{bottom}: One outgoing \emph{orange}  photon starts in the middle.}
\label{tgleich0}
\end{figure}
\begin{figure}[!]
\includegraphics[width=0.49\textwidth]{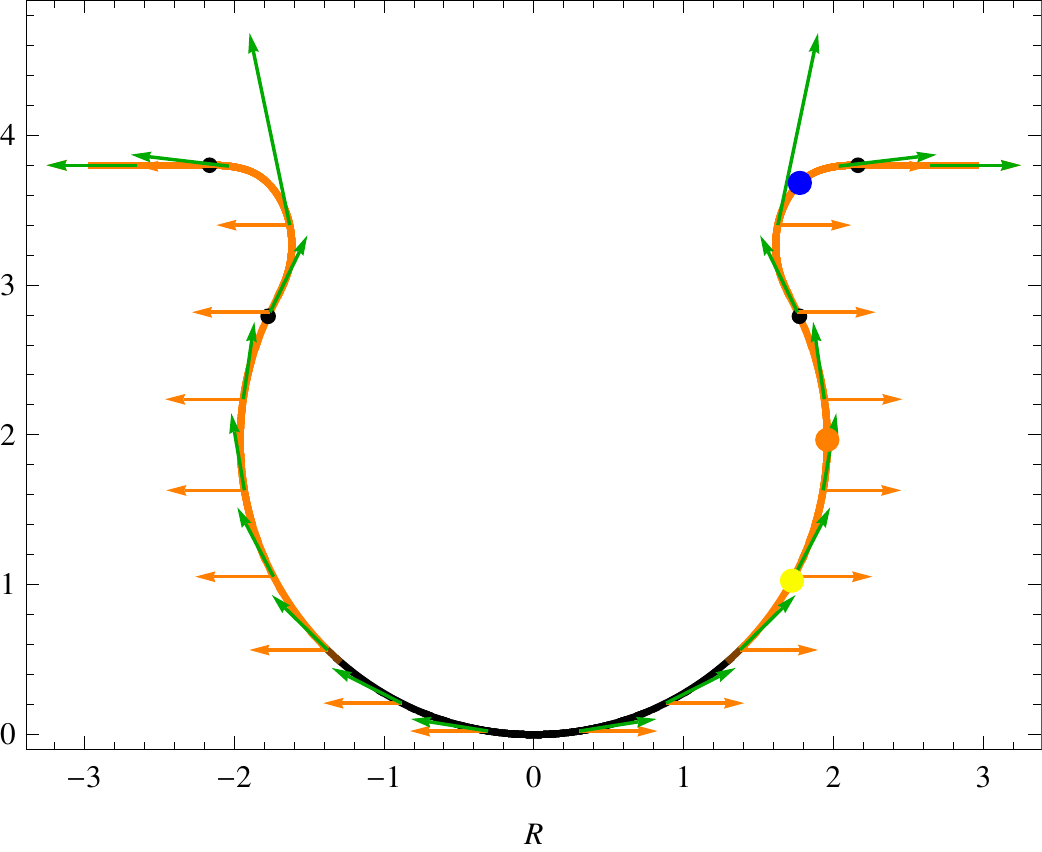}\newline
\includegraphics[width=0.49\textwidth]{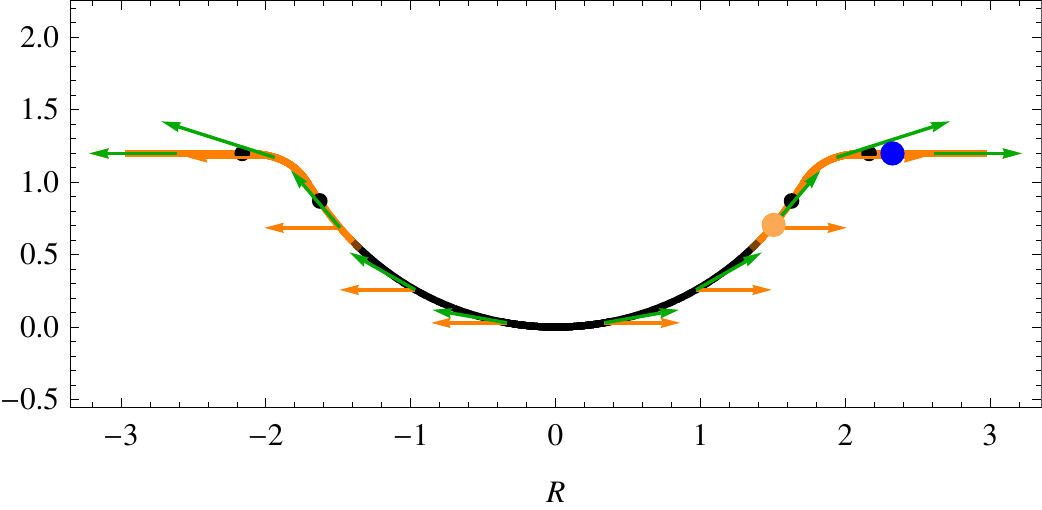} 
  \caption{$t=1.1$\newline
\emph{top}: A third \emph{blue} photon traveling inward appeared. But it is trapped and hence recedes as long as the embedding at its position is \emph{orange}. \newline
\emph{bottom}: The Hubble horizon overtakes the orange outgoing photon.}
\label{tgleich1}
\end{figure}

\begin{figure}[!]
\includegraphics[width=0.49\textwidth]{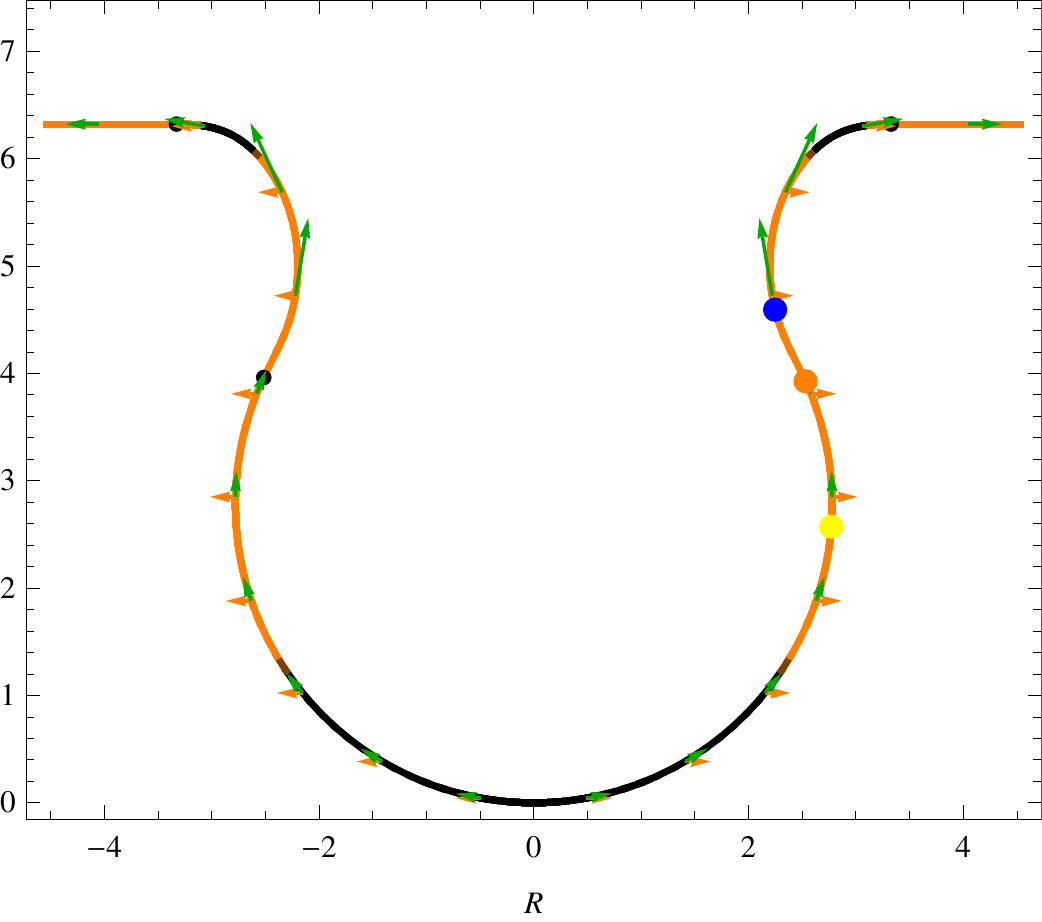}\newline
\includegraphics[width=0.49\textwidth]{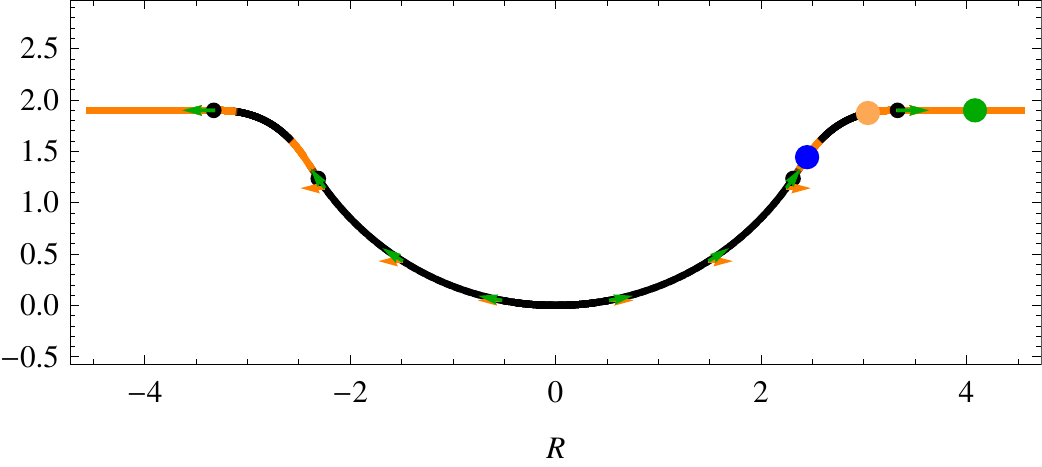}
\caption{$t=2.1$\newline
\emph{top}: Horizon crossing occurred and the neck starts to stretch: the proper distance of the empty region between the \emph{black} dots grows. The extrinsic curvature $K$ gets smaller, but in the neck it remains larger since it has to create the negative intrinsic curvature \ce{^{(3)\!\!}$R$}. \newline
\emph{bottom}:  A third \emph{green} photon traveling inwards appeared.}
\label{tgleich2}
\end{figure}
\begin{figure}[!]
\includegraphics[width=0.49\textwidth]{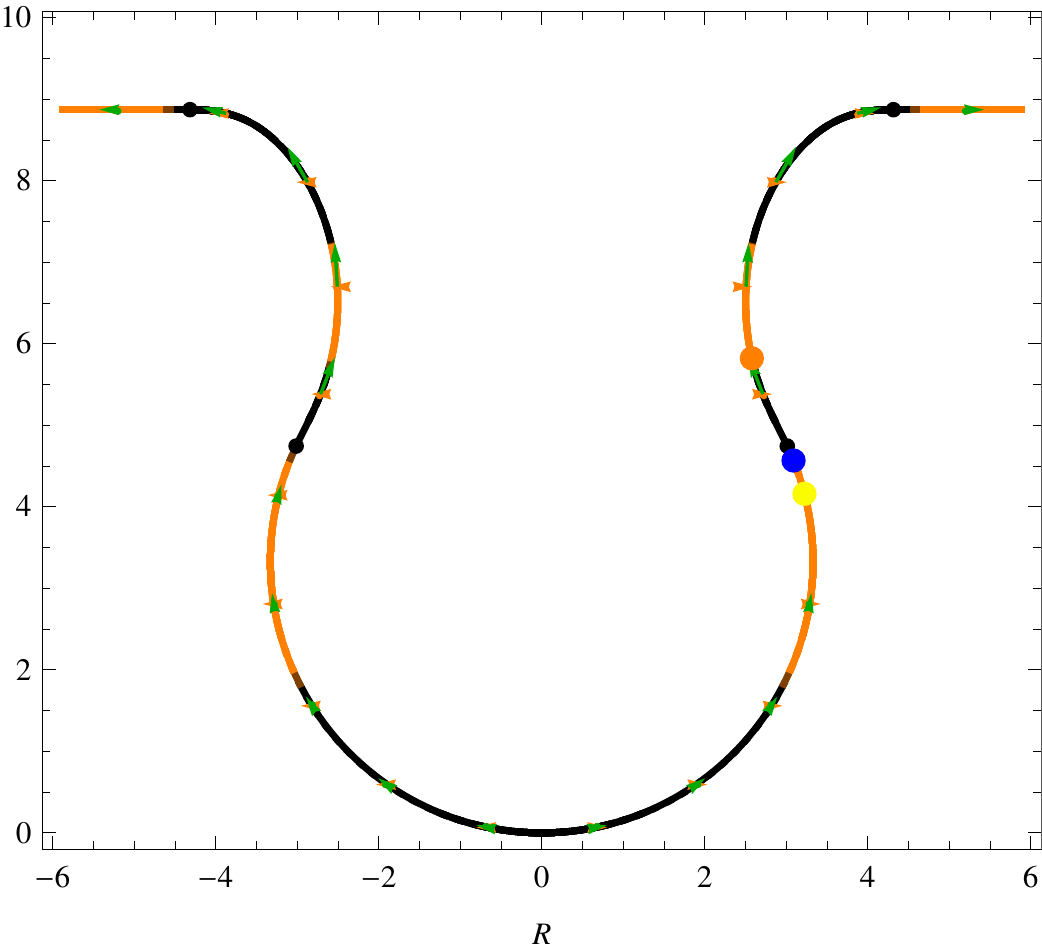}\newline
\includegraphics[width=0.49\textwidth]{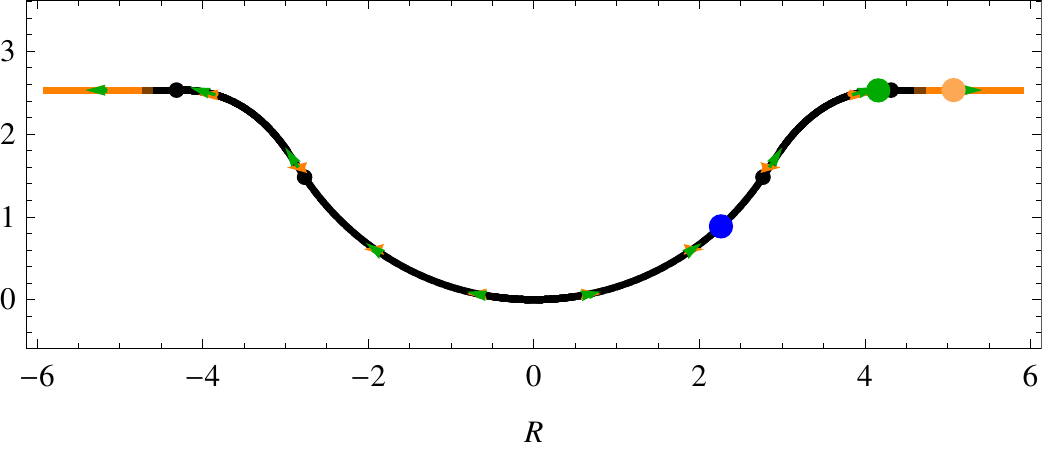}
  \caption{$t=3.1$\newline
\emph{top}: The outer-trapped \emph{orange} region gets fragmented: vertical regions naturally remain trapped for a longer time: Moving in these regions does not alter $R$ much.\newline
\emph{bottom}: The \emph{orange} outgoing photon now overtakes the Hubble horizon.}
\label{tgleich3}
\end{figure}

\begin{figure}[!]
 \includegraphics[width=.49\textwidth]{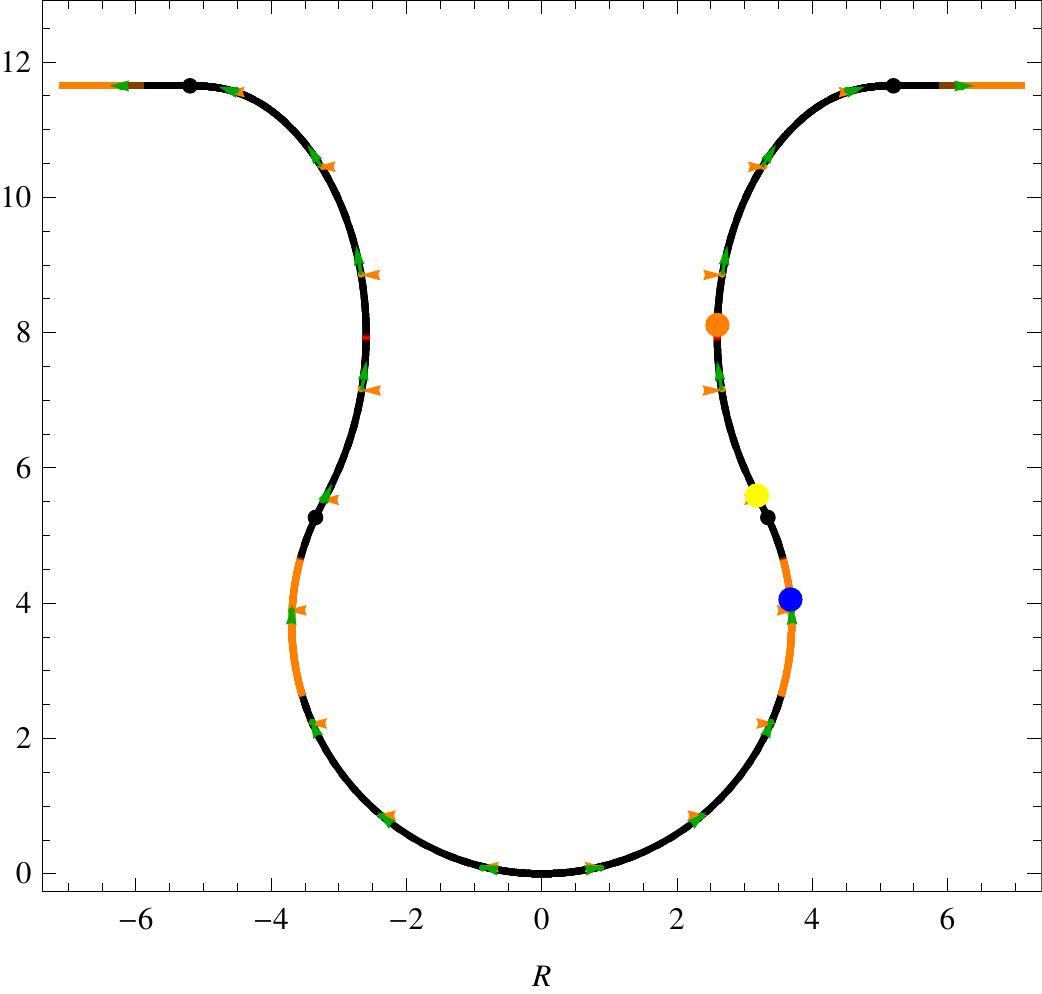}\newline
\includegraphics[width=.49\textwidth]{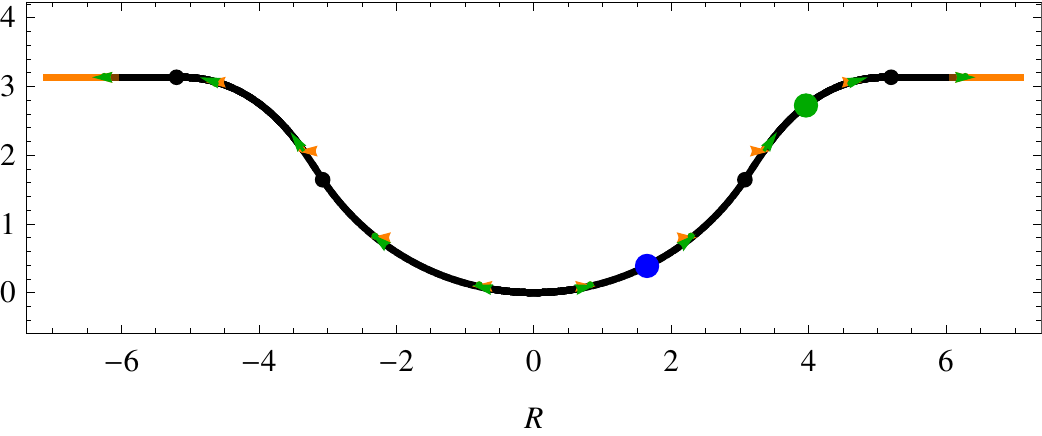}
\caption{$t=4.1$\newline
\emph{top}: Note that type II fluctuations with their local minimal $R$ may be seen as wormholes according to the definition C of \cite{MHC09}.  To prevent a static wormhole from becoming a black hole one needs exotic matter violating a few energy conditions (see \cite{V95}).  In an accelerating universe a dynamical superhorizon-sized wormhole will never reenter the horizon and hence does not require exotic matter to be stable. In our  decelerating dust Universe we would also need exotic matter to stabilize the throat once it entered the horizon. Since we do not have exotic matter we see that just below the \emph{orange} photon at the neck, a tiny \emph{red} future trapped region appeared. This means that the mass of the wormhole got larger than half its Schwarzschild radius. Once this happened the wormhole ceases to be traversable (in the sense that one cannot return) and everything below the outer boundary between \emph{red} and \emph{black} - the black hole apparent horizon -  is doomed.$K^{\!\hat{\chi}}$ in the neck is still positive (\emph{green}) but $H_R$ became negative in the neck center which turned the outer-trapped region into an inner-trapped one (see \eqref{BHhorizon}). Note also that the formation of the trapped region occurred \newline
\emph{bottom}: The \emph{orange} photon leaves the scene. No BH horizon has appeared yet.}
\label{tgleich4}
\end{figure}
\begin{figure}[!]
\includegraphics[width=.49\textwidth]{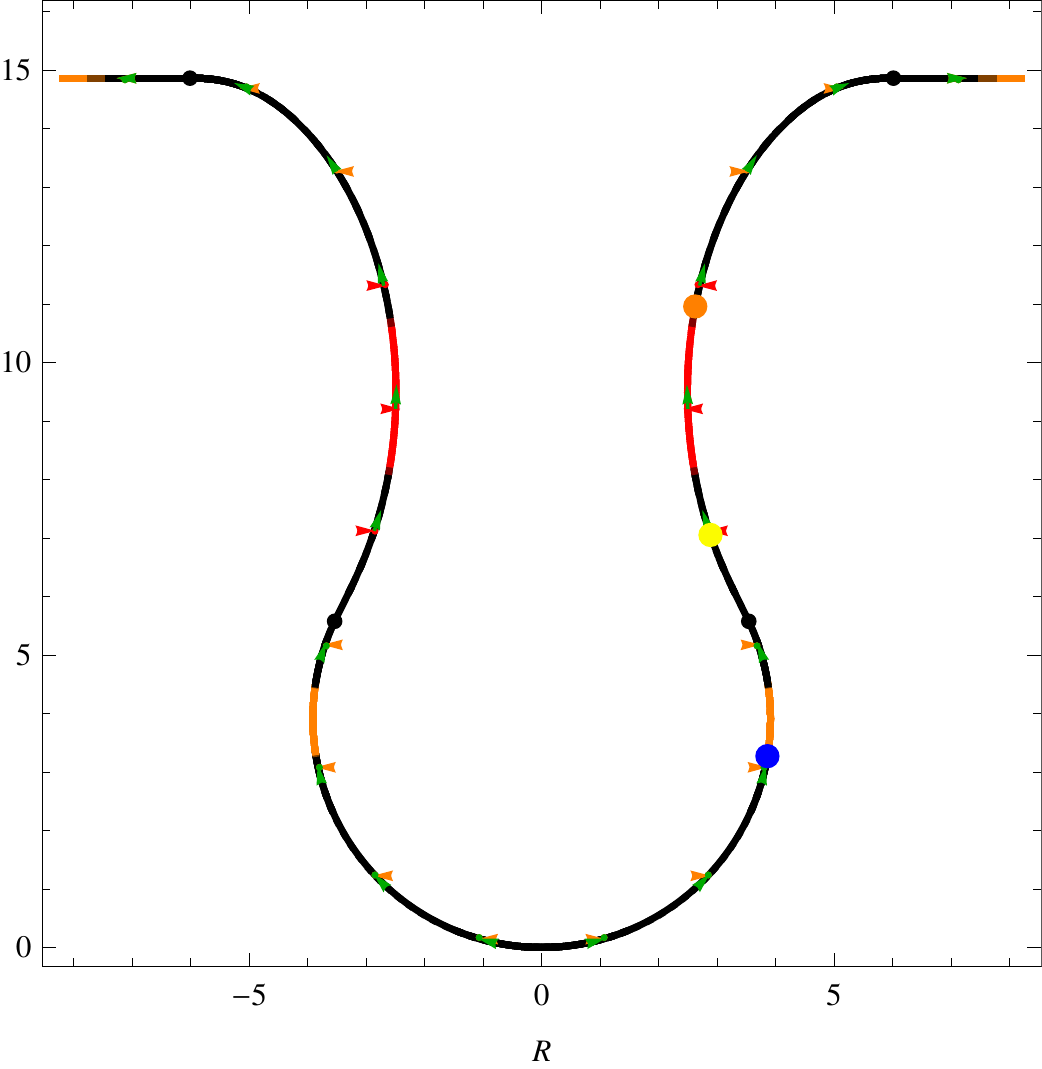}\newline
\includegraphics[width=.49\textwidth]{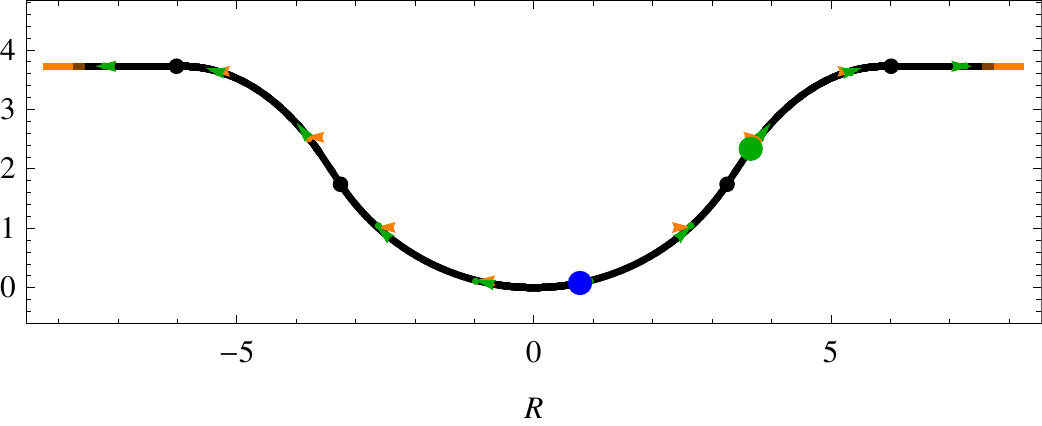}
  \caption{$t=5.1$\newline
\emph{top}: The \emph{orange} photon nearly rests above BH horizon. The trapped contracting \emph{red} region grows due to the stretching of the wormhole driven by the positive $K^{\!\hat{\chi}}$ (\emph{green}). \newline
\emph{bottom}:  The \emph{blue} photon will soon reach the center. }
\label{tgleich5}
\end{figure}

\begin{figure}[!]
 \includegraphics[width=.49\textwidth]{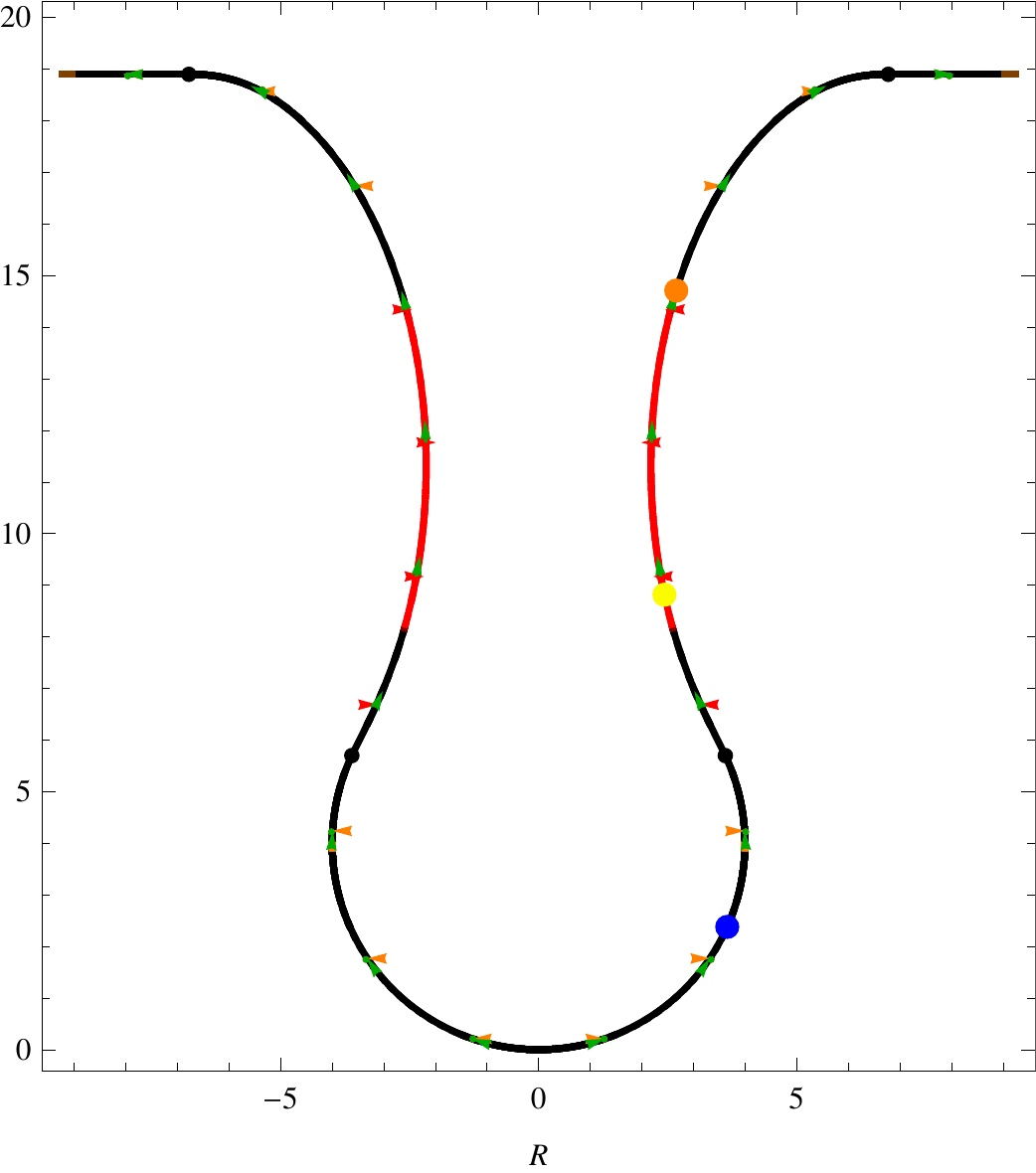}\newline
\includegraphics[width=.49\textwidth]{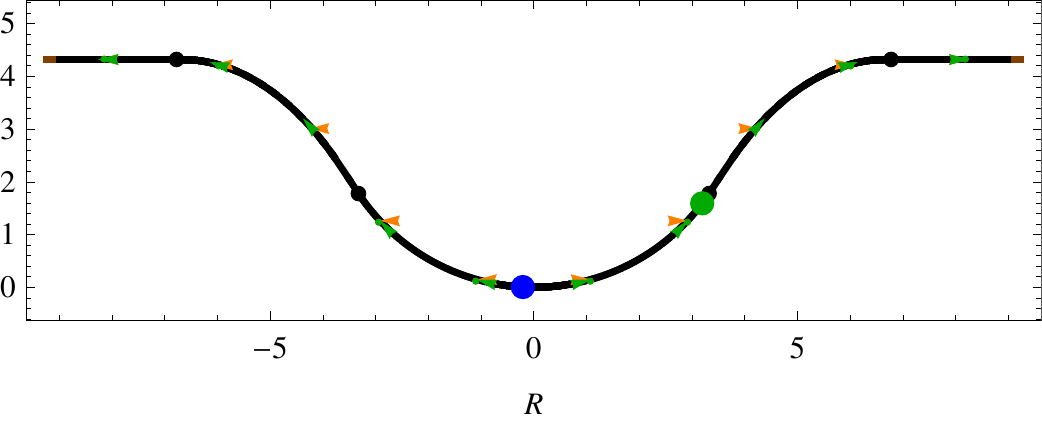}
\caption{$t=6.1$\newline
The visible region is now completely within the Hubble horizon. \newline
\emph{top}: Also the outer-trapped region in the bubble is disappearing. The \emph{blue} and \emph{yellow} photon bravely march towards inevitable oblivion.}
\label{tgleich6}
\end{figure}
\begin{figure}[!]
\includegraphics[width=.49\textwidth]{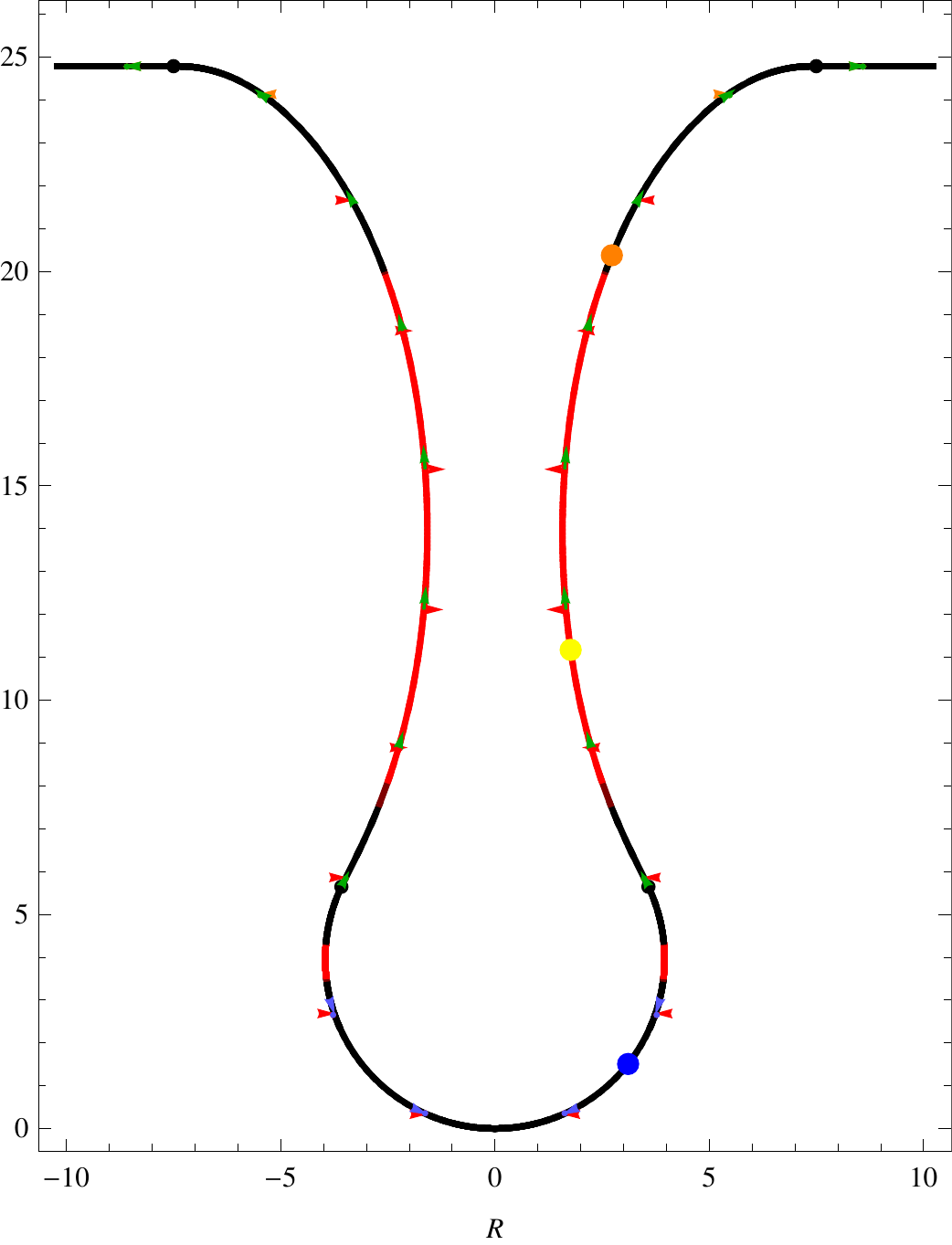}\newline
\includegraphics[width=.49\textwidth]{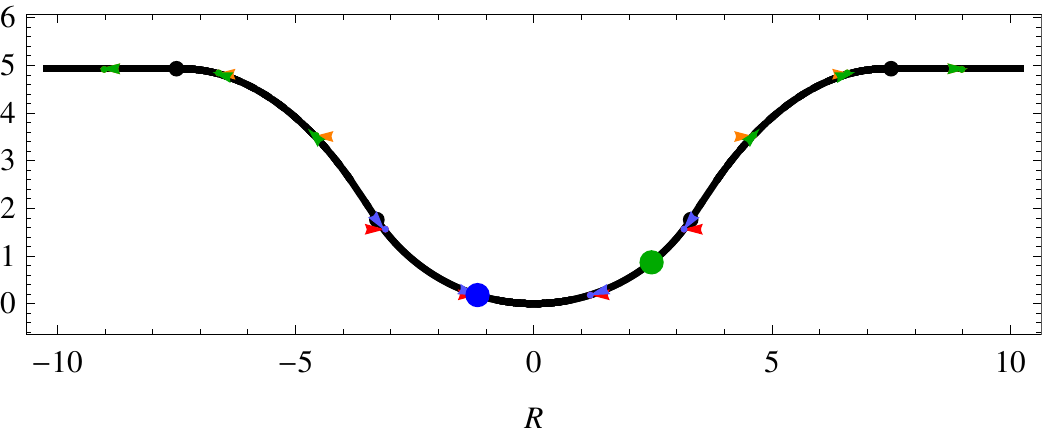}
  \caption{$t=7.1$\newline
\emph{top}: At maximum expansion $t_\mathrm{m}$ the sign of $H_R$ changed in the bubble (\emph{orange} $\rightarrow$ \emph{red}), where a new future trapped horizon has appeared. This is the essential horizon for a type I fluctuation. In the neck  only $H_R$ changed sign and $K^{\!\hat{\chi}}$ remained positive (\emph{green}) such that the second term in \eqref{extriLTB} becomes positive and (visibly) reduces the negative \ce{^{(3)\!\!}$R$} in the neck (see \eqref{hamiconsltb}). This also leads to a direction-dependent contraction/expansion of space. Since in the lower hemisphere $K^{\!\hat{\chi}}$ got negative, space contracts there in every direction.\newline
\emph{bottom}: Here the sign of $H_R$ and $K^{\!\hat{\chi}}$ changed in the whole closed part (which is only part of the lower hemisphere).}
\label{tgleich7}
\end{figure}

\begin{figure}[!]
 \includegraphics[width=.49\textwidth]{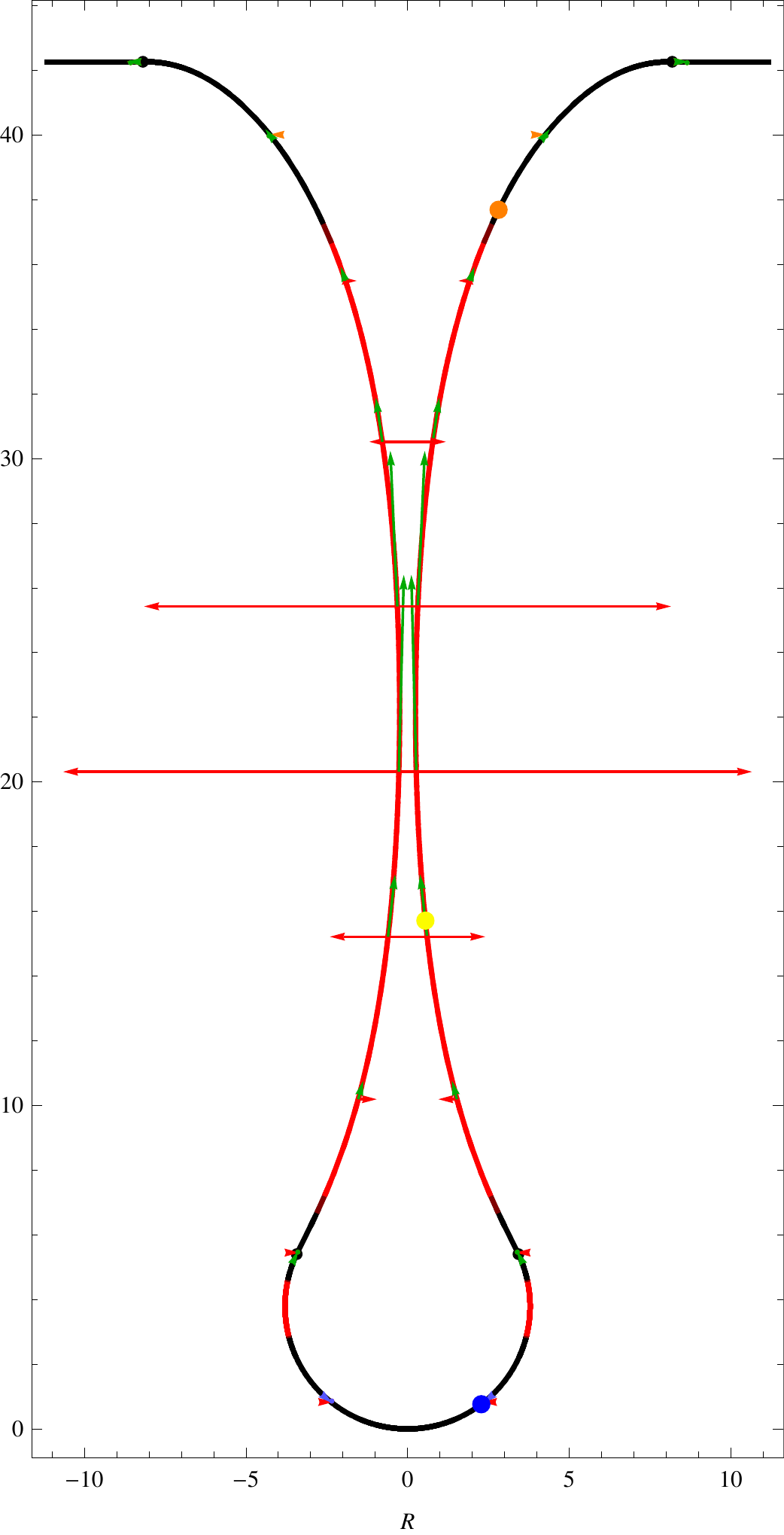}\newline
\includegraphics[width=.49\textwidth]{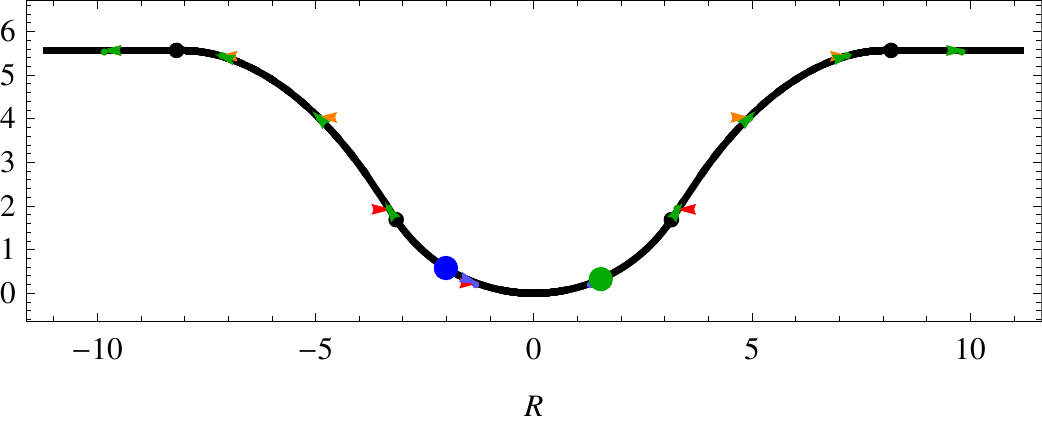}
\vspace*{-5mm}
\caption{$t=8.1$\newline
\emph{top}: In the interior of the wormhole $H_R$ (\emph{horizontal} arrows) is negative   (\emph{red} and inward-directed) everywhere and in the neck center it blows up, while \ce{^{(3)\!\!}$R$} gets smaller since $K^{\!\hat{\chi}}$ (\emph{tangential} arrows) becomes also large but stays positive (\emph{green} and outward-directed). Any matter inside the neck would get extremely spaghettified: radially stretched by the large positive $K^{\!\hat{\chi}}$ but still volume reduced due to the large negative $H_R$. \newline
\emph{bottom}: The \emph{blue} photon marches upwards. Still no sign of a BH horizon. Compare the latitude of the future trapped region in the bubble of the type II fluctuation.}
\label{tgleich8}
\end{figure}
\begin{figure}[!]
\includegraphics[width=.48\textwidth]{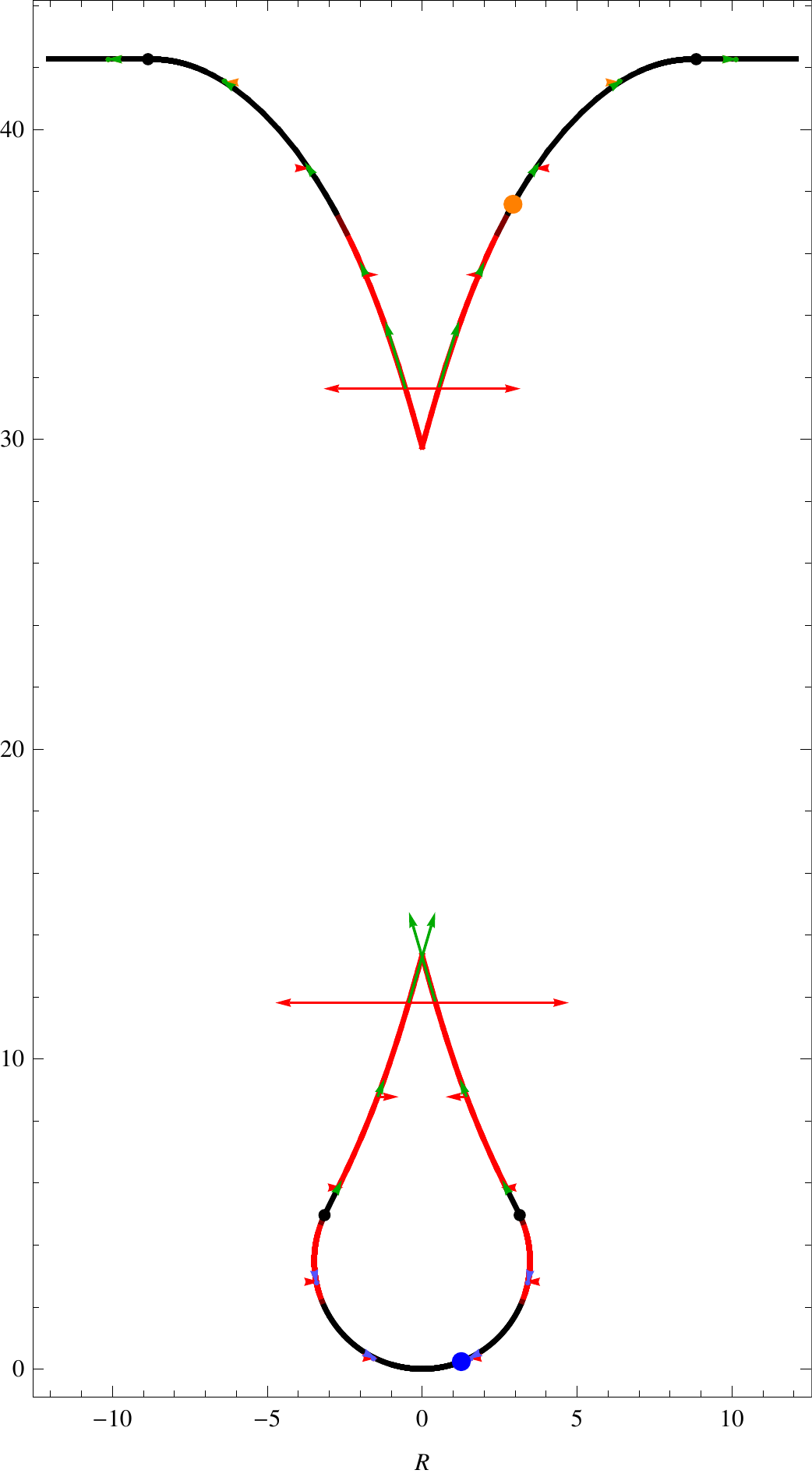}\newline
\includegraphics[width=.48\textwidth]{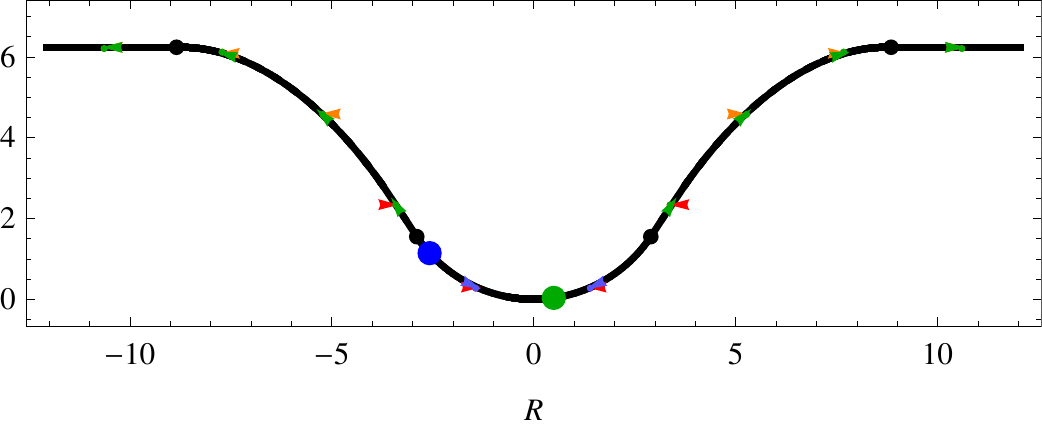}
\vspace*{-10mm}
  \caption{$t=9.1$\newline
\emph{top}: The wormhole has been pinched off: a singularity formed that crushed the \emph{yellow} photon. It is also quite interesting that the singularity formed in a matter free region (between the two \emph{black} dots). Note that the vertical separation of the cusps in the embedding has no physical meaning. The inner-trapped region in Fig.\,\ref{tgleich4} appeared shortly after horizon entry, suggesting that also in the case of a radiation-filled universe the formation of a non-central singularity cannot be avoided: pressure forces will try to bring material trough the wormhole, which is effective only when the throat is within the horizon. Once within it will immediately shrink in the angular direction and stretch in the radial direction. Both processes slow down the pressure driven matter transport. Hence we expect that any type II fluctuation with with $\zeta\geq\zeta_\mathrm{nc}>\zeta_\mathrm{max}$ will form a noncentral singularity, where $\zeta_\mathrm{nc}$ is a new shape dependent threshold.  It would be interesting to determine $\zeta_\mathrm{nc}$ by a proper numerical simulation. \newline
\emph{bottom}: The \emph{green} photon reaches the center.}
\label{tgleich9}
\end{figure}

\begin{figure}[!]
 \includegraphics[width=.49\textwidth]{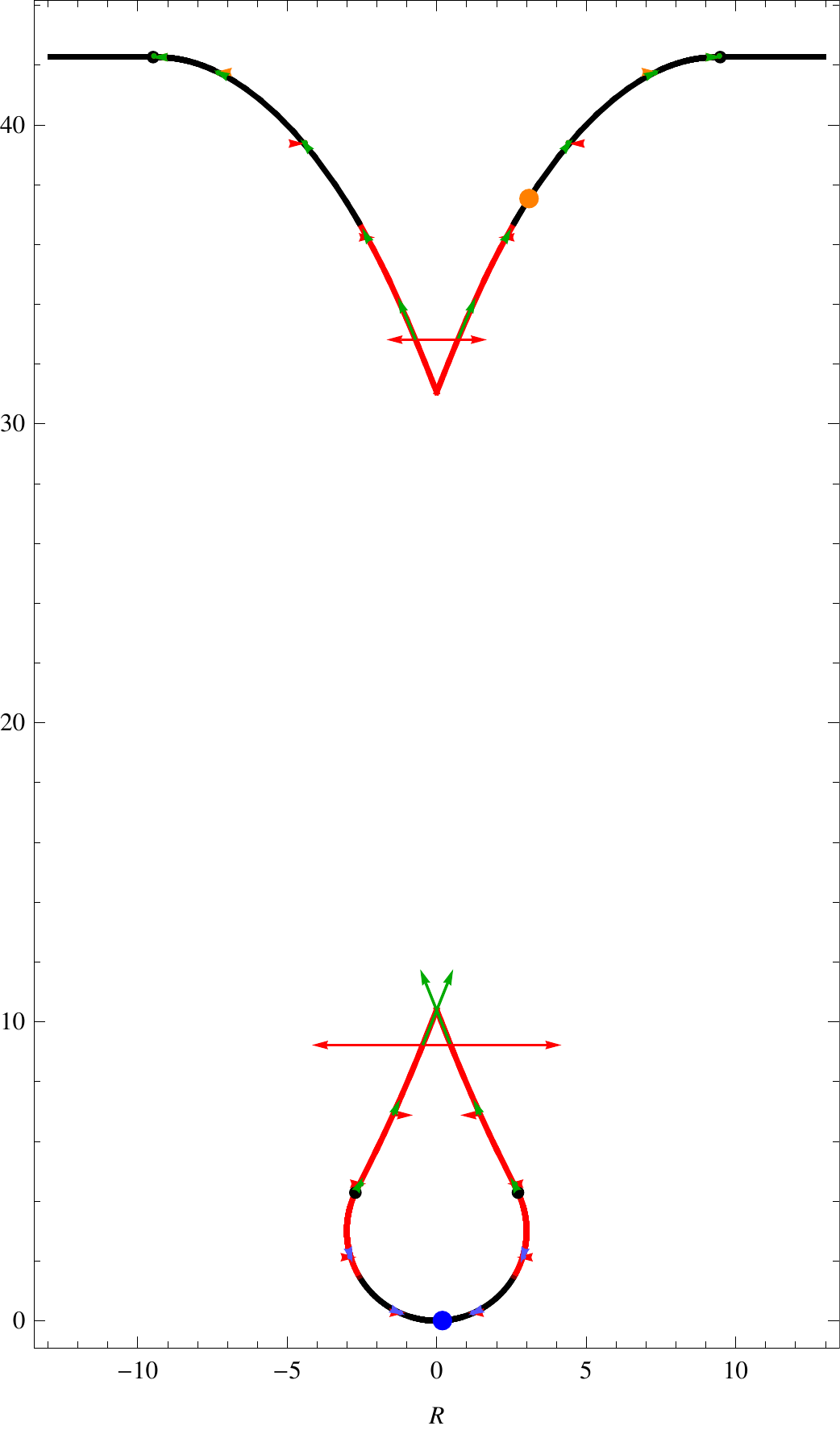}\newline
\includegraphics[width=.49\textwidth]{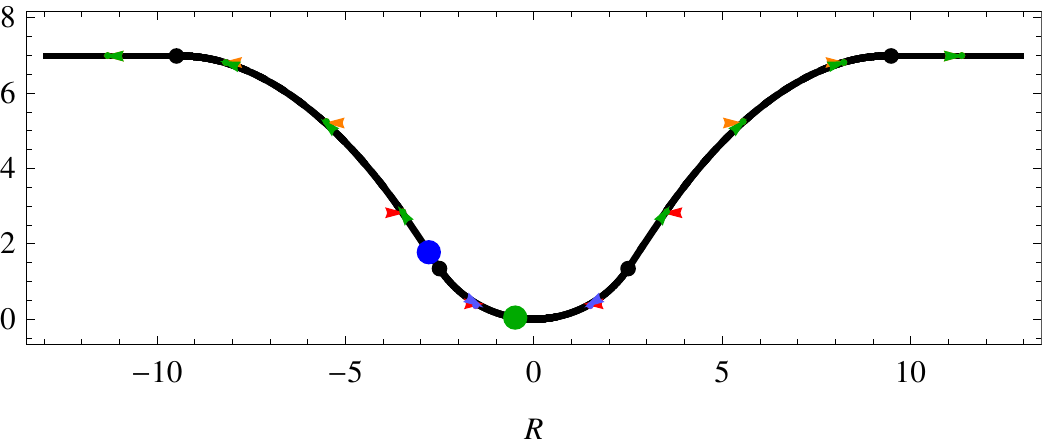}
\caption{$t=10.1$\newline
\emph{top}: The different trapped regions in the baby universe have merged. The \emph{blue} photon reaches the center. \newline
\emph{bottom}: Comparing the latitude $\chi$ of the lower \emph{black} dots with the latitude of the BH horizon in the baby universe in the upper panel we see that the BH horizon will form soon.}
\label{tgleich10}
\end{figure}
\begin{figure}[!]
 \includegraphics[width=.49\textwidth]{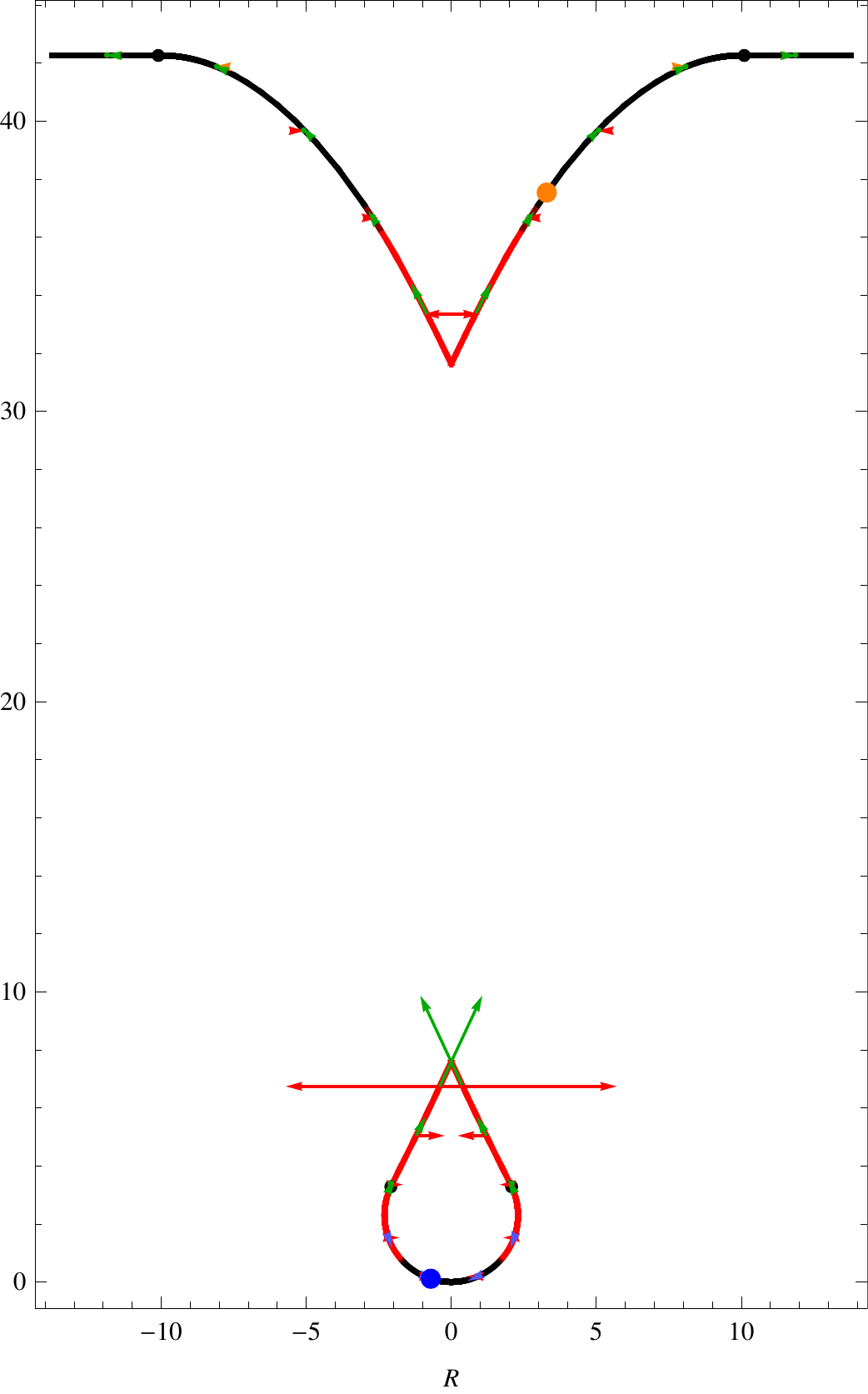}\newline
\includegraphics[width=.49\textwidth]{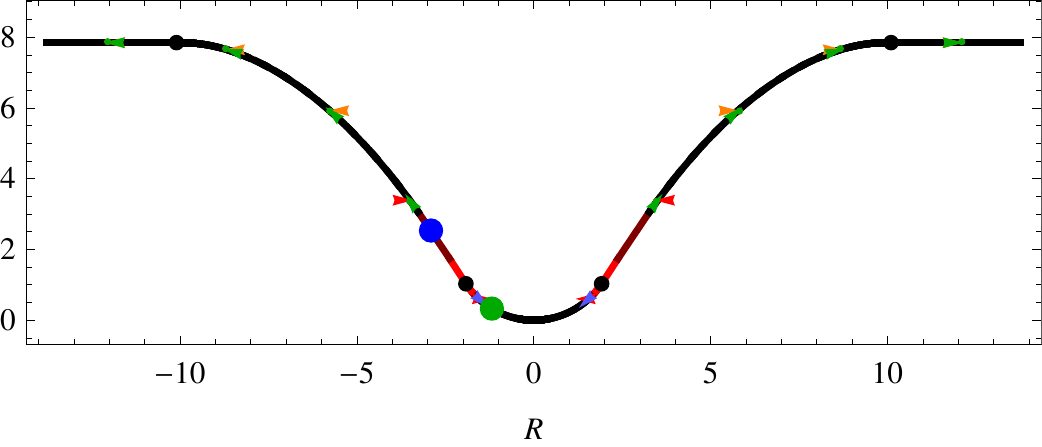}
\caption{$t=11.1$\newline
\emph{top}: The cut-off ``elastic band'' has now relaxed: In the baby universe \ce{^{(3)\!\!}$R$} is  everywhere positive. \newline
\emph{bottom}: The BH horizon has now formed around the \emph{black} mass dots and sealed the fate of the photons and the matter in the overdensity: All the matter and the \emph{green} photon will be crushed. The \emph{blue} photon is lucky. (Here \emph{Mathematica} colored some parts \emph{greyish red}  which actually should be \emph{black}. Hence the \emph{blue} photon is really outside the trapped region. The same holds for Fig.\,\ref{tgleich12})}
\label{tgleich11}
\end{figure}

\begin{figure}[!]
 \includegraphics[width=.49\textwidth]{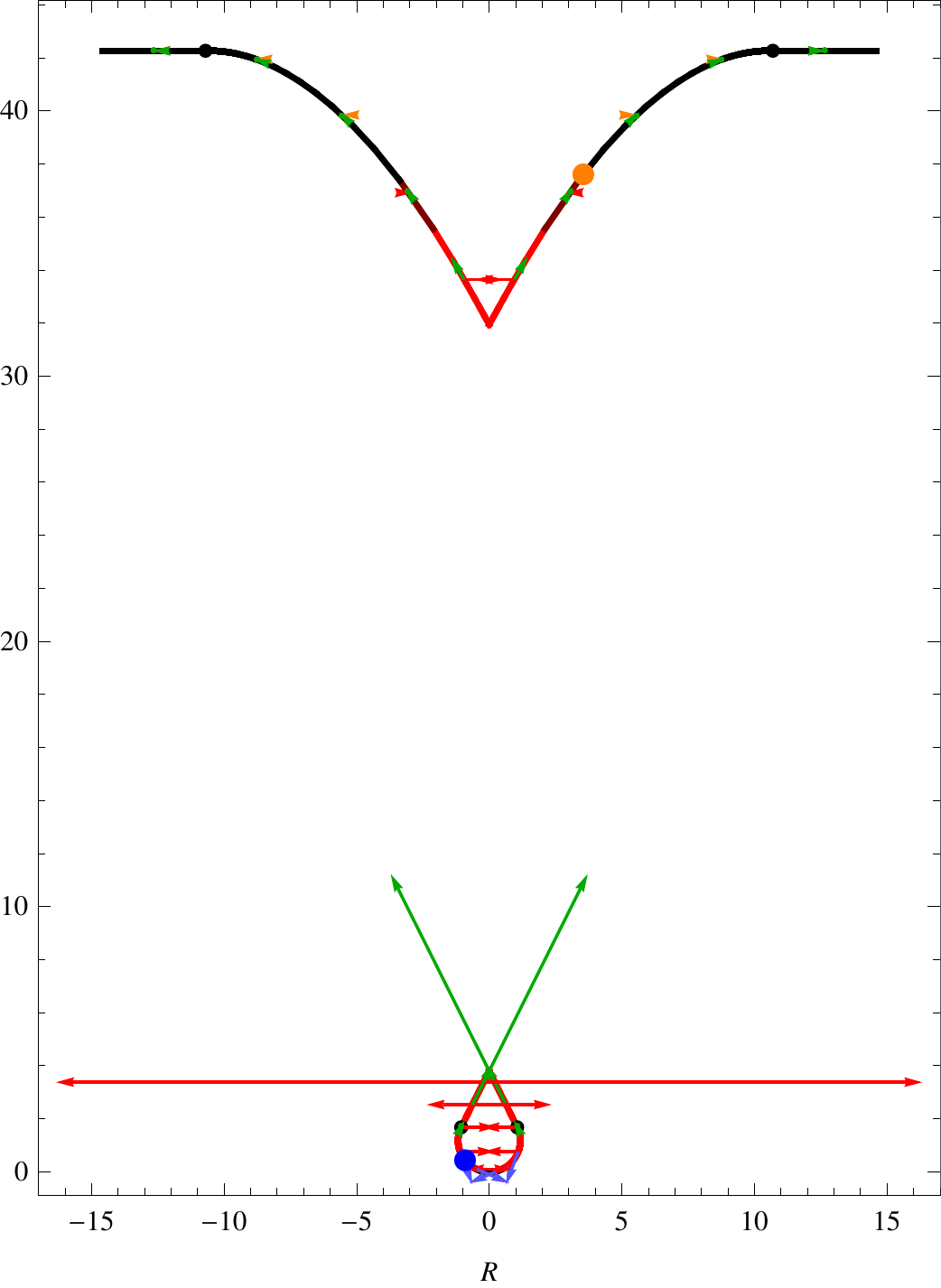}\newline
\includegraphics[width=.49\textwidth]{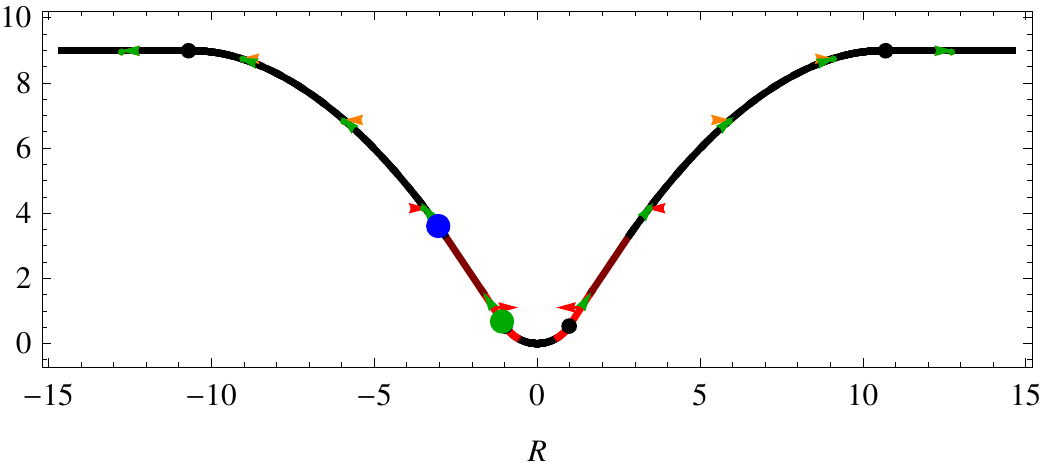}
\caption{$t=12.1$\newline
\emph{top}: The \emph{orange} photon still nearly rests above the horizon. The baby universe is so close to its big crunch that the \emph{blue} photon will not reach the PBH singularity in the baby universe. \newline
\emph{bottom}: Note that the singularity in the type II fluctuation in Fig.\,\ref{tgleich8} was announced by a blowup of $K$ (while \ce{^{(3)\!\!}$R$} got smaller). Here it is \ce{^{(3)\!\!}$R$} that blows up while $K$ remains small.}
\label{tgleich12}
\end{figure}

\begin{figure}[!]
 \includegraphics[width=.49\textwidth]{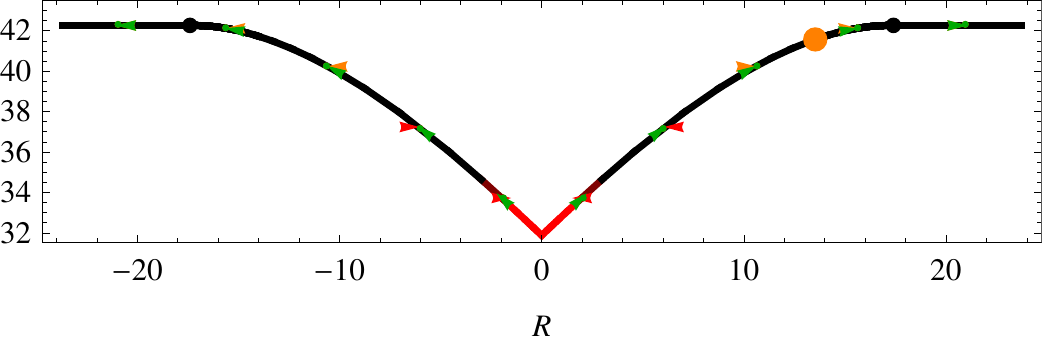}\newline
\includegraphics[width=.49\textwidth]{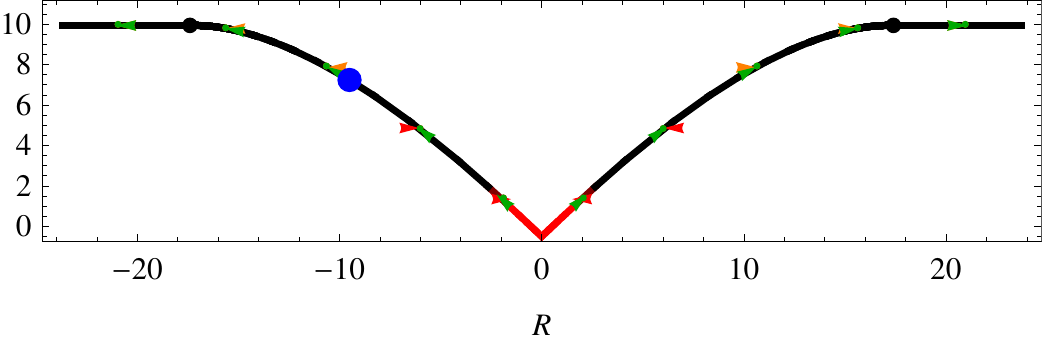}
\caption{$t=25.1$\newline
Both photons required a long time (twice the age of the universe after PBH formation at $t_\bh=12.57$) to finally escape the horizon, although they are still inside the matter free vacuole: they must be strongly redshifted. After $t_\bh$ both slices are now physically identical and indistinguishable: a PBH with Schwarzschild radius $R_\bh=2.6$ and mass $M_\bh=R_\bh/2$ inside an expanding vacuole embedded in a flat exact FRW universe. The metric inside the comoving \emph{black} dots is exactly Schwarzschild such that the existence of the FRW is undetectable using gravity experiments like Kepler law or peri-PBH-shift inside the vacuole. This has been proven by Einstein and Strauss \cite{ES45}.}
\label{tgleich25}
\end{figure}
\FloatBarrier
\subsection{Conformal diagrams}
\begin{figure}[t]
\includegraphics[width=0.43\textwidth]{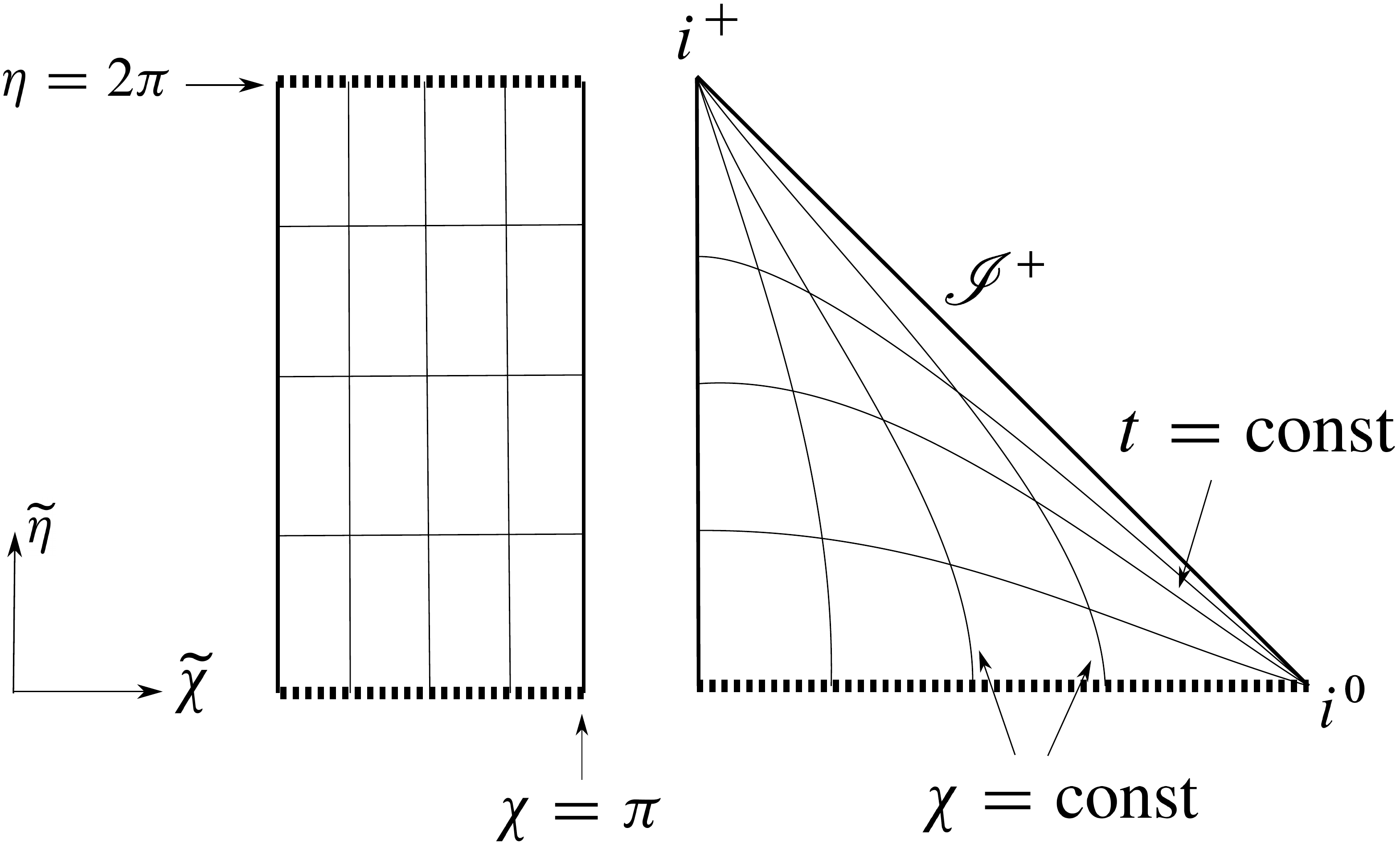}
\caption{Conformal diagrams for closed (\emph{left}) and flat (\emph{right}) dust FRW.}
\label{confopenclose}
\end{figure}

Despite the great insight into geometrodynamics that embeddings can give us, the causal structure of spherically symmetric spacetimes is best visualized in the form conformal diagrams \cite{M05}, where one maps the whole spacetime with a finite coordinate system ${\tilde{\eta},\tilde{\chi}}$ with the property that lightlike radial geodesics are straight lines with:
$$\tilde{\chi}(\tilde{\eta})=\pm \tilde{\eta}+\mathrm{const}.$$
\FloatBarrier
The conformal diagrams for flat and closed dust FRWs are depicted in Fig.\, \ref{confopenclose}.

The right diagram shows a flat FRW spacetime. The diagonal boundary $\mathscr{I}^+$ is the future null infinity, where all photons go (they are reflected at the vertical $\chi=0$ line). All spatial slices end for $\chi\rightarrow\infty$ in $i^0$, called spacelike infinity, while all timelike trajectories terminate for $t\rightarrow\infty$ in the future timelike infinity  $i^+$. The horizontal dashed line is the big bang singularity. The left diagram showing a closed dust FRW has a simpler structure, also the original comoving coordinates $\eta,\chi$ are already conformal coordinates ${\tilde{\eta},\tilde{\chi}}$. Photons are reflected at $\chi=0$ and $\chi=\pi$ lines, originate in the lower big bang singularity and terminate in the upper big crunch singularity.

Before we look at the conformal diagrams of our two spacetimes let us first discuss the seeming separation of the baby universe seen in the embedding diagrams of the type II fluctuation.
One might have wondered whether the formation of the baby universe corresponds to a topology change of the spatial slice which is proved to be impossible (see \cite{HE73,V95}).
Note that even a ``normal'' dust collapse in asymptotic Minkowski space which looks quite like a type I collapse, can be foliated in such an unnatural way to look like a type II (see \cite{MTW73} box 23.1). But here we did not change the slicing! So where does this change in the behavior of $t$=const slices come from?

\begin{figure}[!]
\includegraphics[width=0.4\textwidth]{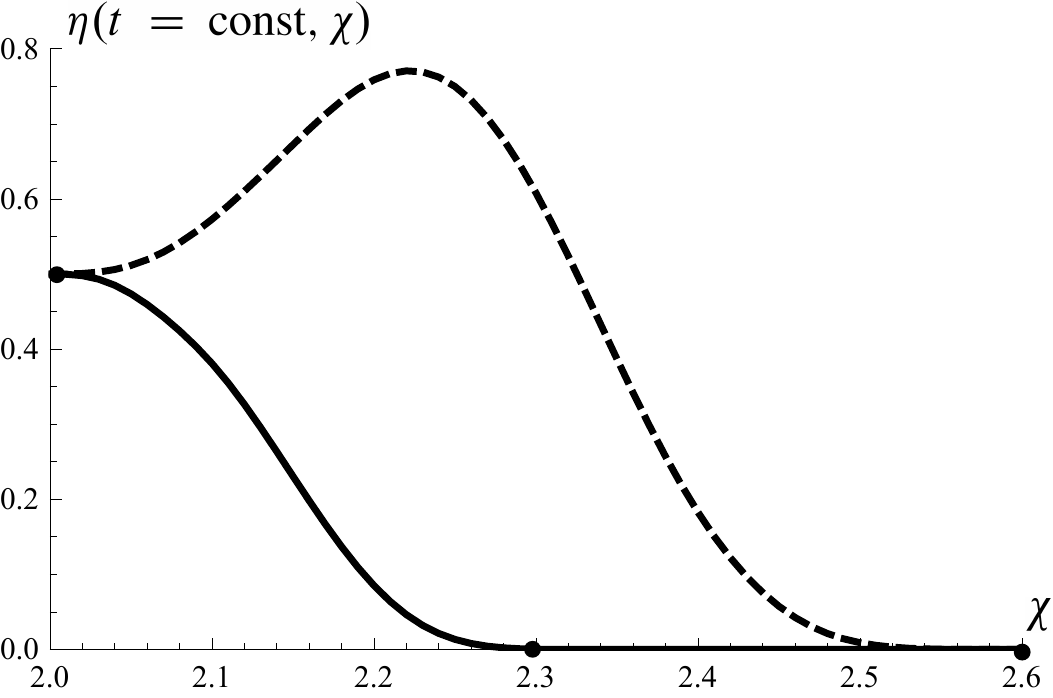}
\caption{Spatial shape of $\eta$ in the transition region. Type II is plotted \emph{dashed}, type I \emph{full} (shifted for better comparison)}
\label{etavonchi}
\end{figure}

\begin{figure*}[!]
\flushleft
\includegraphics[width=0.99\textwidth]{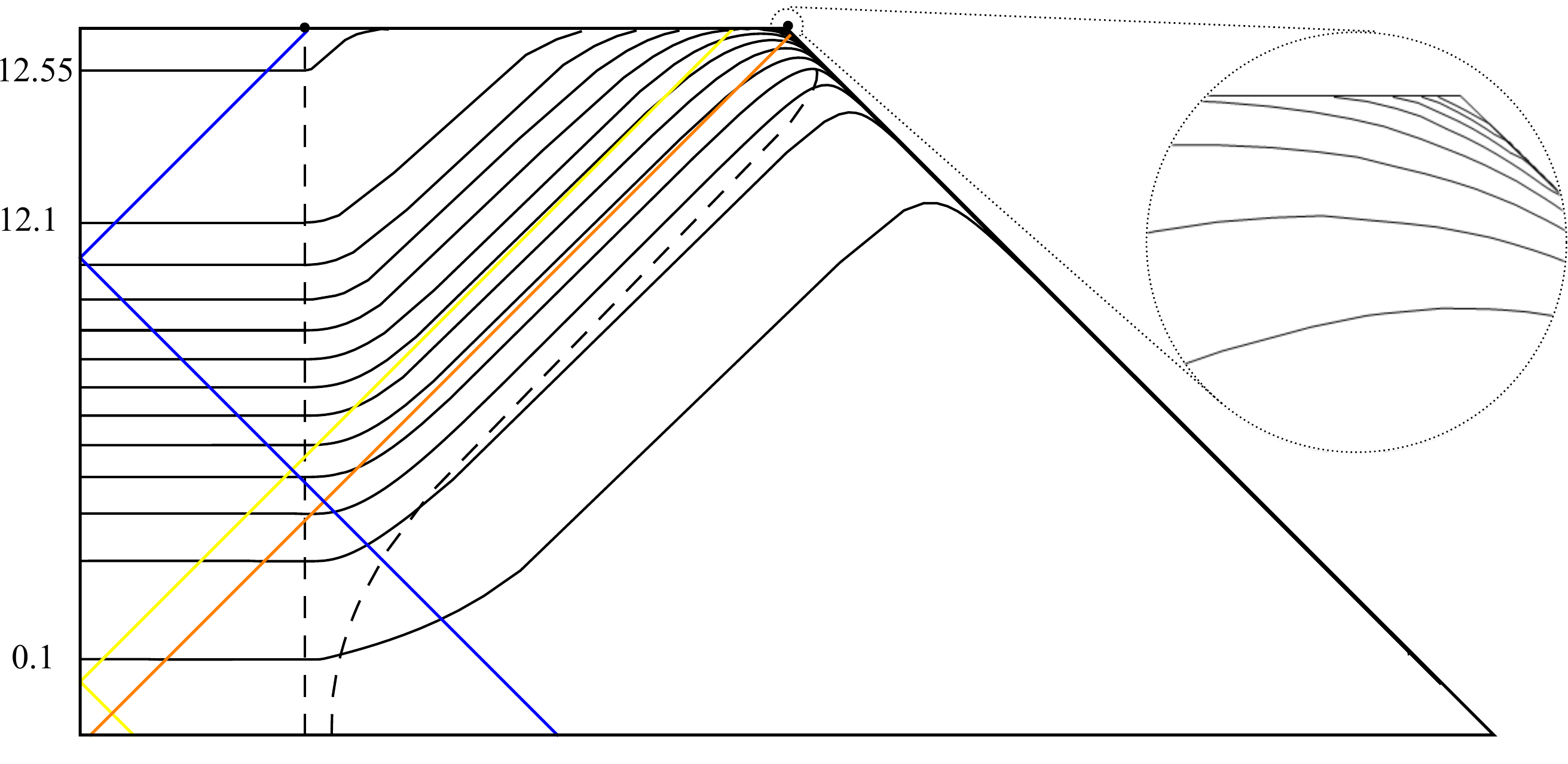}
\includegraphics[width=0.8\textwidth]{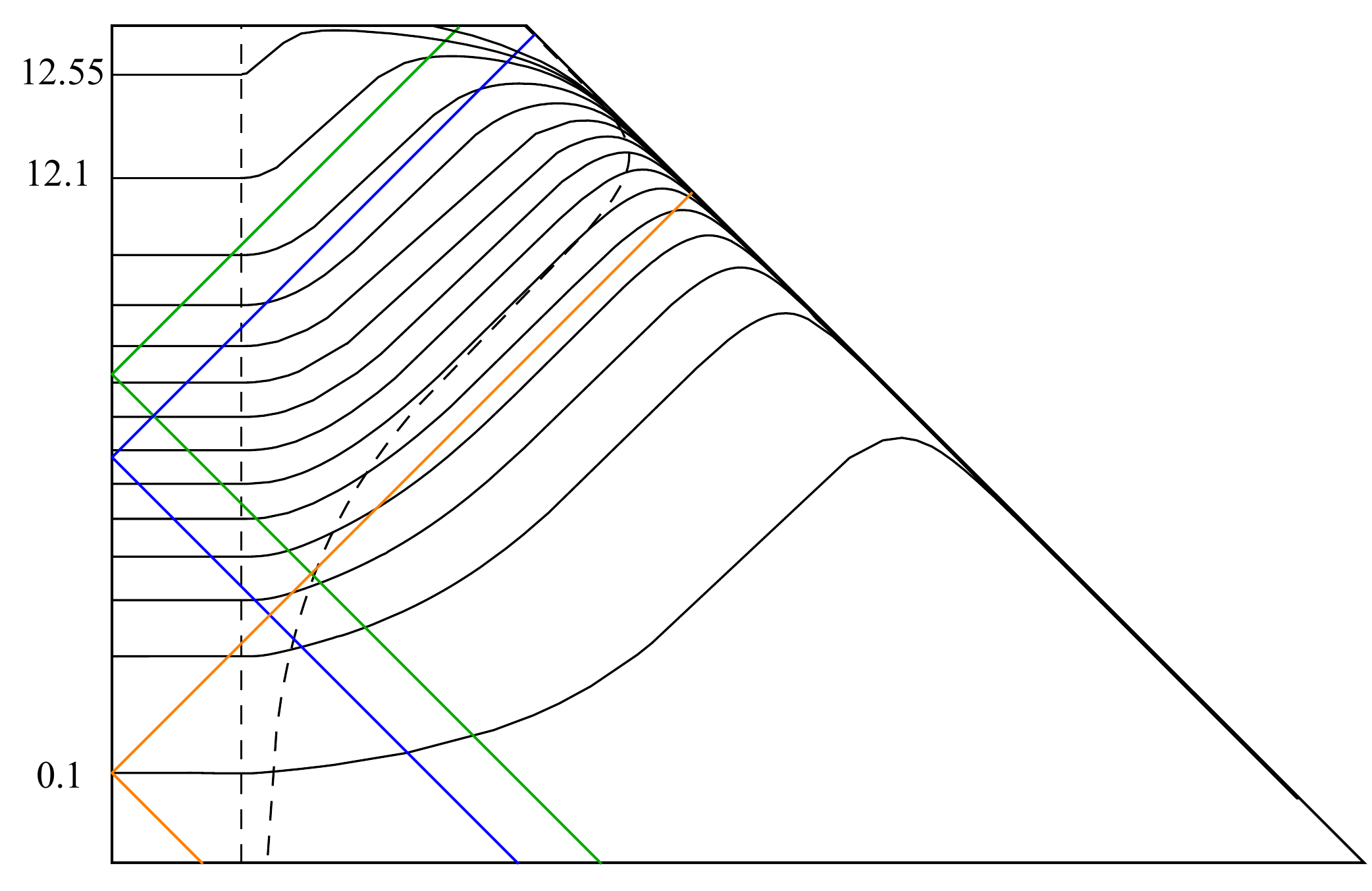}
\caption{Conformal diagrams of type II (\emph{top}) and I (\emph{bottom}) spacetime. The coordinates $\tilde{\eta},\tilde{\chi}$ are chosen to keep $\tilde{\eta}=\eta,\,\tilde{\chi}=\chi$ in the exact closed FRW part and to have a $\tilde{\eta}$-simultaneous big bang and big crunch. The \emph{full} spacelike lines are $t=\,$const slices and the ones from $t=0.1$ and to$t=12.1$ coincide with the slices shown in the embeddings. Close before the big crunch in the exact closed FRW at $\tilde{\eta}= 2\pi$ corresponding to $t=12.56$, we added an additional line at $t=12.55$. Above this in the conformal diagram for the type I fluctuation the slice at $t=13.1$ terminating in the singularity can be seen, showing that the \emph{green} photon still lives at this time. The magnified corner of the type II diagram shows the termination of $t>8.15$-slices in the singularity coming from the flat FRW. They reappear from the singularity well visibly on the left. This effect corresponds to the seeming topology change we have seen in the embedding diagrams of the type II fluctuation. For the type I fluctuation spatial slices coming from the flat FRW only end in the singularity but do not reappear for smaller $\tilde{\chi}$. If we had used the $\tilde{\eta}$-slicing in the embeddings, the evolution of type II would have looked very similar to the one of type I. Hence the type I and II spacetimes are very similar. The two \emph{dashed} timelike lines correspond to the comoving boundaries that separate the exact closed and flat FRW parts from the matching region. They are depicted as \emph{black} dots in the embeddings. The \emph{straight, diagonal} (\emph{colored}) null lines correspond to the \emph{large} (\emph{colored}) photon dots in the embeddings. Note that in the type I diagram the upper part of the \emph{dark} (\emph{blue}) photon line is approximately the BH event horizon, while for the type II spacetime it is a photon line starting in the left corner which is approximately the \emph{orange} photon.}
\label{confopencloseII}
\end{figure*}

Let us calculate the spatial shape of $\eta(t\!=\!\mathrm{const},\chi)$ in the transition region for both fluctuations. We can easily solve \eqref{LTBsol} for $\eta$ which gives
$$\eta(t,\chi)=\frac{|E|^{3/2}}{M}t+\sqrt{2\frac{R|E|}{M}-\left(\frac{RE}{M}\right)^2}.$$
The main $\chi$ dependence is in the first term. Figure \ref{etavonchi} shows it.
The common feature of type I and II is that $\eta(\chi)$ interpolates between a constant finite $\eta$ in the closed FRW part and 0 in the flat part (this ensures that there is no big crunch).
We see that the $t$=const-slice of the type II fluctuation has a maximum while the type I slice does not. Since the mass function $M$ was chosen to be constant in the transition region the whole effect comes from $|E|^{3/2}$. The fact that $|E_\tii|$ exhibits a maximum at $\chi=2.2$ while $|E_\ti|$ does not, explains the strange difference between type I and II seen in the embedding diagrams.
Note that the appearance of a local maximum of $|E_\tii|=|\Phi_\tii`^2-1|$ in the transition region, is inevitable and not just an artifact of the chosen matching, since $\Phi_\tii`$ has to change its sign somewhere in the matching region (compare Fig.\,\ref{massenergy}). This means that in synchronous gauge every arbitrary type II fluctuation (which may be defined in general by the nonmonotonicity of $R(\chi)$) has this appearance of the collapse.
Since our model fluctuation spacetimes consist of parts of the spacetimes depicted in Fig.\,\ref{confopenclose} smoothly glued together, the shapes of the resulting conformal diagrams for type I and II are obvious. A numerical construction of such diagrams from the same data used for the embeddings is shown in Fig.\,\ref{confopencloseII}.
Most important is that type I and type II have the same conformal structure. From the conformal diagrams we see: the $t$=const-slices in the transition region cause all the difference we saw in the embedding diagrams. While for type I fluctuations the big crunch is first reached in the closed FRW part and later in the matching region, for type II fluctuations the big crunch is first reached in the matching region and later on in the closed FRW part. And there is no apparent topology change if one uses the $\tilde{\eta}$-slicing. 

Although the embeddings show only the case of a spherically symmetric dust collapse, it  is clear that a slightly aspherical type II fluctuation in a radiation filled universe will also feature a non-central singularity formation in any reasonable slicing and provided that $\zeta\geq\zeta_\mathrm{nc}$ (see Fig.\,\ref{tgleich8}). The only possibility to prevent this for $\zeta<\zeta_\mathrm{nc}$ is the  time dependence of $E$, such that the neck could widen through the pressure once the fluctuation is inside the horizon and turn it into a type I shortly before PBH formation. If this does not happen, then the PBH mass should be approximately given by Eq.\,\eqref{bhmass}. However only a numerical simulation can give a definite answer.
\FloatBarrier

%\clearpage
\section{Conclusion}
\label{concl}

Type II ($\zeta>\zeta_\ma$) and marginal ($\delta=\delta_\ma$) fluctuations sometimes misinterpreted as separate universes do form primordial black holes.

The formation process looks completely different compared to the type I class ($\zeta<\zeta_\ma$) in synchronous gauge.
This is only a gauge artifact of the synchronous and related ``natural'' slicings and can be avoided in the quite unnatural $\tilde{\eta}$-slicing used in the conformal diagram.
%Another artifact of the synchronous slicing is the existence of a maximum density fluctuation $\delta_\ma$ which corresponds to a fluctuation that marks the boundary between type I and II.
%The reason for this is twofold: (\emph{i}) In a radiation filled universe every overdensity has at maximal expansion a density contrast of exactly 4. (\emph{ii}) The earlier a fluctuation enters the horizon before maximal expansion the smaller will be the density contrast at horizon crossing which defines $\delta$. Since this time span is minimal for $\delta_\ma$ fluctuations because of the spatial geometry, they have the maximum $\delta$. 
The appearance of $\delta_\ma$ is independent of the no-SU condition formulated by Carr and Hawking \eqref{rhomaxhaw}. It is impossible to violate this condition due to \eqref{rhomaxhar}. The actual limiting value to form a SU obtained by closing a type II is $\delta_\su\rightarrow 0$ corresponding to $\zeta\rightarrow\infty$.
The nonmonotonicity of $\delta(\zeta)$ suggests not to use the density fluctuation variable $\delta$ but instead the curvature fluctuation variable $\zeta$ if one also wants to consider larger fluctuations, for instance in a probability distribution function of fluctuation amplitudes.

For homogeneous overdensities every type I fluctuation has a $\delta$-twin in the type II class related by $\chi_\ti=\pi-\chi_\tii$. If there is no pressure, both fluctuations form a PBH with exactly the same mass although the formation process looks very different for a type II. This is related to the fact that the gravitating mass energy that one can probe by using Kepler's law is not the integral of the density over the proper volume but over the volume of the would-be flat space using the areal radius $R$ (see Eq.\,\eqref{MSmass}).
A difference between type I and II arises if there is pressure present. We expect the existence of a shape and pressure dependent threshold $\zeta_\mathrm{nc}$ with the property that for $\zeta\geq\zeta_\mathrm{nc}>\zeta_\mathrm{max}$ a non-central singularity forms, while for $\zeta_\mathrm{nc}>\zeta\geq \zeta_\mathrm{max}$ the type II fluctuation relaxes via pressure driven matter loss within the horizon to a centrally collapsing fluctuation. To find this threshold it is necessary to perform a numerical simulation.
\appendix
\section{derivation of the curvature fluctuation}
\label{serfluc}
For this we rewrite the spatial metric
\begin{equation}
\mathrm{\ce{^{(3)\!\!}$ds^2$}} = \frac{R'^2}{1+E} \md r^2  + R^2 \md\Omega^2 \nonumber
\end{equation}
in terms of the radial approximately comoving  $\chi$-coordinate: 
$$\chi(r)=\int_0^r \frac{\md r'}{\sqrt{1+E(r')}}.$$
Denoting the inverse by $\Phi(\chi)=r$, $R(\Phi(\chi),t)\equiv R(\chi,t)$ and a derivative w.r.t. $\chi$ by $`$ we get:
\begin{equation}
\mathrm{\ce{^{(3)\!\!}$ds^2$}} = \frac{{R`^2}(\chi,t)}{\Phi`^2(\chi)} \md \chi^2  + R^2(\chi,t) \md\Omega^2.\label{ltbchi}
\end{equation}
From the definition of $\chi$, we obtain the relation:
$$E(\chi)=\Phi`^2(\chi)-1.$$
Writing the metric in terms of the 3d-conformally flat and noncomoving radial coordinate $s$:
\begin{equation}
\mathrm{\ce{^{(3)\!\!}$ds^2$}} = c^2(t) e^{2 \zeta(s,t)} \left(\md s^2  +   s^2 \md\Omega^2\right) \label{ltbs}
\end{equation}
we obtain from \eqref{ltbchi} and \eqref{ltbs} these two equations:
\begin{equation}
    \frac{{R`}}{\Phi`}\md \chi= c e^{\zeta} \md s\qquad R=c e^{\zeta} s .\label{coortrafoltb}
\end{equation}
Taking the quotient gives
\begin{equation}
    \frac{{R`}}{R \Phi`} \md \chi = d \ln s. \label{confs}
\end{equation}
Let us assume again that the fluctuation is localized such that for $\chi\geq \chi_b$ space is flat: $E(\chi)=0$ or $\Phi`(\chi)=1$ implying $\Phi(\chi)=\chi-\chi_b+\Phi(\chi_b)$. 
Flatness also requires a spatially constant $\zeta$. Choosing $\zeta=0$ for $s \geq s_b=s(\chi_b)$ allows us to set $r=s$ hence $\Phi(\chi)=s$ (in the flat space) and we get from \eqref{coortrafoltb} $R`=b=c$.
Inside the fluctuation we integrate \eqref{confs} from $\chi$ to $\chi_b$ which gives $s(\chi,t)$:
\begin{equation}
    s(\chi,t)=\Phi(\chi_b) \exp\left(-\int_{\chi}^{\chi_b}\frac{{R`}}{R \Phi`} \md \chi ' \right).
\end{equation}
Inserting this result into the second equation of \eqref{coortrafoltb} we obtain $\zeta(s(\chi,t),t)$
\begin{equation}
    \zeta(\chi,t)= \ln \frac{R}{R`(\chi_b)\Phi(\chi_b)}+\int_{\chi}^{\chi_b}\frac{{R`}}{R \Phi`} \md \chi '. \label{zetaformori}
\end{equation}
In the case of a homogeneous overdensity patched to flat space at $\chi_b \simeq \chi_a$ we have $\Phi(\chi)=\sin \chi$, $R=a \sin \chi$ and $R`=a$ at $\chi_a$:
\begin{equation}
    \zeta(\chi,t)\simeq \ln \frac{\sin \chi}{\sin(\chi_a)}+\int_{\chi}^{\chi_a}\frac{\md \chi '}{\sin \chi '} .
\end{equation}
In fact, this converges for $\chi \rightarrow 0$ and we obtain:
\begin{equation}
    \zeta(0,t)\equiv\zeta= -2\ln \cos \frac{\chi_a}{2} .
\end{equation}
Note that $\chi_b \simeq \chi_a$ (cf. Fig.\,\ref{sphconfig}) is only a good approximation as long as the perturbation exceeds the horizon. Otherwise gradients in the matching region between $\chi_a$ and $\chi_b$ become important and hence $E$ and $\chi_a$ and $\chi_b$ themselves become time dependent.

\section{derivation of the density fluctuation}
\label{derden}
In order to obtain the ratio $\Delta_\hc=\rho/\bar{\rho}$ at $t_{\mathrm{hc}}$ we use the time dependence of these two densities determined by the Friedmann Eq.\,\eqref{friedmann} of a flat and closed radiation-filled FRW, respectively.
Using $t_\mathrm{m}=a_\mathrm{m}$, we compute the ratio at $t_{\mathrm{hc}}$:
$$\frac{\rho_{\mathrm{hc}}}{\bar{\rho}_{\mathrm{hc}}}=\frac{4 (t_\mathrm{hc}/t_\mathrm{m})^2}{\sin^4 \eta_{\mathrm{hc}}}\equiv\frac{4 v^2}{\sin^4 \eta_{\mathrm{hc}}}.$$
The condition for horizon crossing is $R_\mathrm{hc}=H^{-1}_\mathrm{hc}$ or
$$a_\mathrm{m} \sin \eta_{\mathrm{hc}} \sin \chi_a = 2 t_{\mathrm{hc}}\quad\Rightarrow\quad  \sin \eta_{\mathrm{hc}} \sin \chi_a = 2 v.$$
Using \eqref{conftime} to express $\eta$ in terms of $t$, we obtain
$$\sqrt{2 v-v^2} \sin \chi_a = 2 v.$$
The solution is
$$v=\frac{2 }{4\sin^{-2}\chi_a+ 1}$$
which we use to obtain
\begin{eqnarray}
\Delta(t_\mathrm{hc})&=&\nonumber\frac{4 v^2}{\sin^4 \eta_{\mathrm{hc}}}=\frac{4 v^2}{\left(2 v / \sin \chi_a  \right)^4}=\frac{\sin^4 \chi_a }{4v^2}\\\nonumber
&=& \left(1+\frac{1}{4} \sin^{2}\chi_a\right)^2
\end{eqnarray}
and finally $\delta$ becomes
\begin{equation}
\delta=\frac{1}{16}\sin^{2}\chi_a\left( 8+ \sin^{2}\chi_a \right).
\end{equation}
\section{Conditions on $E$ and $M$}
\label{LTBcons}
In the LTB spacetime the functions $M(r)$ and $E(r)$ can be chosen as initial conditions on a given spacelike slice $t=\mathrm{const}$ in order to model a desired $\rho$ or spatial geometry. $E$ and $M$ cannot be chosen freely: 
\begin{enumerate}
\item[({\it i})] $E$ and $M$ must be sufficiently smooth since they enter Einstein equations and at $r=0$ we need from Eq.\,\eqref{hamicons}, e.g. $E\sim r^2$ and $M\sim r^3$, since naturally $\dot{R}^2\sim r^2$ near $r=0$.
\item[({\it ii})] $E\geq -1$ to keep the spacetime Lorentzian, hence $R'=0$ where $E=-1$.\\
\item[({\it iii})] $\rho\geq 0$ everywhere, e.g. from Eq.\,\eqref{massenergy} $M'\leq 0$ whenever $R'<0$.
\item[({\it iv})] There is no shell crossing: shells with larger $r$ must not overtake shells with smaller $r$ since at those points $R'=0$ and hence $\rho$ diverges. 
\item[({\it v})] It is important to note that $R'=0$ is not forbidden but requires at this point $M'=0$. In order to keep the metric Lorentzian at points where $R'=0$ we need there additionally $E=-1$. 
\end{enumerate}
Point ({\it iv}) does not have that much physical significance. We made an overidealisation: spherical symmetry and noninteracting particles. In reality shocks will form when the calculation says shell crossing and the solution cannot be trusted once a shell crossing has occurred.  Note that shell crossing even arises for perfect fluids \cite{LL06}, implying that not the missing pressure is the problem but rather the spherical symmetry or the perfect fluid description.
%Unfortunately condition ({\it iv}) requires the strongest finetuning on the functions $E$ and $M$. But this finetuning has only the practical reason that we can trust our idealised (but simple) calculation and it does not in general mean that shell-crossing free initial conditions given by $E$ and $M$ are extraordinarily special: The appearance of a shock during gravitational collapse will complicate the BH formation (delaying/preventing the collapse of the overdense region or parts of it) and so choosing no-shell-crossing conditions just corresponds to choosing initial conditions which lead to BH formation without complications, which then allows us to use the (over-) idealised LTB setup for dust. 
For a detailed treatment of no-shell-crossing conditions see \cite{HL85}, and for a complete treatment of points ({\it i - v}) and alternatives see \cite{MH01}.
\newline

\acknowledgements 
The authors would like to thank Bernard Carr, Slava Mukhanov, Kristina Giesel, Felix Berkhahn, Andreas Joseph and Stephen Appleby for helpful comments and inspiring discussion. MK is grateful to Andreas Fackler for finding the conformal diagram construction algorithm. The work of SH and JW was supported by the TRR 33 'The Dark Universe'.
\bibliography{bhrdm}

%-------------------------------------------------------------------------------The End-----------------------------------------------------------------

\end{document}